\documentclass[journal]{IEEEtran}

\usepackage{myStyleIEEE}
\usepackage{nomencl}
\usepackage{ragged2e}
\usepackage[thinc]{esdiff}
\IEEEoverridecommandlockouts

\newlength{\bracewidth}

\usepackage{etoolbox}
\renewcommand\nomgroup[1]{%
  \item[\bfseries
  \ifstrequal{#1}{P}{A. Parameters}{%
  \ifstrequal{#1}{V}{C. Variables}{%
  \ifstrequal{#1}{S}{B. Sets and Indices}{}}}%
]}

\begin{document}

\title{Coordinated Planning for Stability Enhancement in High IBR-Penetrated Systems}

\newtheorem{proposition}{Proposition}
\renewcommand{\theenumi}{\alph{enumi}}

\author{Zhongda~Chu,~\IEEEmembership{Member,~IEEE,} and
        Fei~Teng,~\IEEEmembership{Senior Member,~IEEE} 
        
        
\vspace{-0.5cm}}
\maketitle
\IEEEpeerreviewmaketitle

\begin{abstract}
Security and stability challenges in future power systems with high penetration Inverter-Based Resources (IBR) have been anticipated as \textcolor{black}{one of the main barriers} to \textcolor{black}{decarbonization}. Grid-following IBRs may become unstable under small disturbances in weak grids, while during transient processes, system stability and protection may be jeopardized due to the lack of sufficient Short-Circuit Current (SCC). To solve these challenges and achieve decarbonization, the future system has to be carefully planned. However, it remains unclear how both small-signal and transient stabilities can be considered during the system planning stage. \textcolor{black}{In this context, this paper proposes a coordinated planning model of different resources in the transmission system, namely the synchronous condensers and GFM IBRs to enhance system stability.} The system strength and SCC constraints are analytically derived by considering the different characteristics of synchronous units and IBRs, which are further effectively linearized through a novel data-driven approach, where an active sampling method is proposed to generate a representative data set. The significant economic value of the proposed coordinated planning framework in both system asset investment and system operation is demonstrated through detailed case studies.
\end{abstract}

\begin{IEEEkeywords}
system planning, synchronous condenser, energy storage systems, short-circuit current
\end{IEEEkeywords}

\makenomenclature
\renewcommand{\nomname}{\textcolor{black}{Nomenclature}}
\mbox{}
\nomenclature[S]{$g\in\mathcal{G}$}{set of SGs}
\nomenclature[S]{$s\in\mathcal{S}$}{set of SCs}
\nomenclature[S]{$c\in\mathcal{C}$}{set of IBRs}
\nomenclature[S]{$c\in\mathcal{C}_l$}{set of GFL IBRs}
\nomenclature[S]{$c\in\mathcal{C}_m$}{set of GFM IBRs}
\nomenclature[S]{$n\in\mathcal{N}$}{set of nodes in scenario tree}
\nomenclature[S]{$t\in T$}{set of time steps in the planning problem}
\nomenclature[S]{$\omega\in \Omega$}{set of linearization data}
\nomenclature[S]{$\Omega_{1,2,3}$}{subsets of $\Omega$}
\nomenclature[S]{$\Delta\Omega$}{data set increment}
\nomenclature[S]{$m$}{iteration index for active sampling}
\nomenclature[S]{$ij \in \mathcal{R}$}{set of branches}

\nomenclature[P]{$p_{\mathrm{min},g}$}{minimum power output of SG $g$}
\nomenclature[P]{$p_{\mathrm{max},g}$}{maximum power output of SG $g$}
\nomenclature[P]{$T^{mut}_g$}{minimum up time of SG $g$}
\nomenclature[P]{$T^{mut}_g$}{minimum up time of SG $g$}
\nomenclature[P]{$T^{st}_g$}{start-up time of SG $g$}
\nomenclature[P]{$\nu$}{parameter for linearization}
\nomenclature[P]{$E_g$}{internal emf of SG}
\nomenclature[P]{$E_s$}{internal emf of SC}
\nomenclature[P]{$X_g$}{reactance of SG}
\nomenclature[P]{$X_s$}{reactance of SC}
\nomenclature[P]{$I_g$}{equivalent current source of SG}
\nomenclature[P]{$I_s$}{equivalent current source of SC}
\nomenclature[P]{$Y_g$}{susceptance of SG}
\nomenclature[P]{$Y_s$}{susceptance of SC}
\nomenclature[P]{$d_0$}{reactive current droop gain of IBR $c$ based on its own capacity}
\nomenclature[P]{$S_g$}{capacity of SG}
\nomenclature[P]{$S_{\mathrm{base}}$}{System base power}
\nomenclature[P]{$\underline{S}_s$}{lower capacity limit of SC}
\nomenclature[P]{$\Bar{S}_s$}{upper capacity limit of SC}
\nomenclature[P]{$\Bar{N}_s$}{maximum number of SCs to be installed}
\nomenclature[P]{$\beta$}{temporary overloading factor}
\nomenclature[P]{$\underline{S}_{c_m}$}{lower limit of $S_{c_m}$}
\nomenclature[P]{$\Bar{S}_{c_m}$}{upper limit of $S_{c_m}$}
\nomenclature[P]{$\Bar{N}_{c_m}$}{maximum number of GFM IBR to be installed}
\nomenclature[P]{$\mathrm{gSCR}_{\mathrm{lim}}$}{gSCR limit}
\nomenclature[P]{$c^{s}$}{annualized investment cost coefficient of SC}
\nomenclature[P]{$\pi_n$}{probability of scenario $n\in \mathcal{N}$}
\nomenclature[P]{$c_g^{st}$}{cost coefficient for startup cost}
\nomenclature[P]{$c_g^{nl}$}{cost coefficient for no-load cost}
\nomenclature[P]{$c_g^{m}$}{cost coefficient for marginal cost}
\nomenclature[P]{$c^{VOLL}$}{value of lost load}
\nomenclature[P]{$Y^0$}{admittance matrix of the transmission lines}
\nomenclature[P]{$\mathcal{K}'$}{linear coefficient for SCC constraints}
\nomenclature[P]{$\mathcal{K}$}{linear coefficient for gSCR constraints}
\nomenclature[P]{$\bar P_{c_m}^{\mathrm{ch}}$}{upper bound of the BESS charging rate}
\nomenclature[P]{$\bar P_{c_m}^{\mathrm{dch}}$}{upper bound of the BESS discharging rate}
\nomenclature[P]{$\eta_{c_m}$}{BESS charging efficiency}
\nomenclature[P]{$\mathrm{SoC}_{\mathrm{max}}$}{upper bound of BESS state of charge}
\nomenclature[P]{$\mathrm{SoC}_{\mathrm{min}}$}{lower bound of BESS state of charge}

\nomenclature[V]{$I_F$}{short-circuit current at Bus $F$}
\nomenclature[V]{$V_F(0)$}{pre-fault voltage at Bus $F$}
\nomenclature[V]{$Z$}{system impedance matrix}
\nomenclature[V]{$\Phi(c)$}{mapping from IBR $c$ to the corresponding bus index}
\nomenclature[V]{$I_{fc}$}{fault current from IBR $c$}
\nomenclature[V]{$I_{Lc}$}{pre-fault load current from IBR $c$}
\nomenclature[V]{$V_{\Phi(c)}$}{post-fault voltage at bus $\Phi(c)$}
\nomenclature[V]{$V_{\Phi(c)}(0)$}{pre-fault voltage at bus $\Phi(c)$}
\nomenclature[V]{$\Delta V_{\mathcal{C}}$}{vector of IBR terminal voltage deviations}
\nomenclature[V]{$Z_{F\mathcal{C}}$}{vector collecting $Z_{F\Phi(c)},\,\forall c\in\mathcal{C}$}
\nomenclature[V]{$I_{F_{\mathrm{lim}}}$}{current limit at Bus $F$}
\nomenclature[V]{$I_c^{\mathrm{max}}$}{maximum current injection of IBR $c$}
\nomenclature[V]{$x_g$}{binary decision variable for SG online state}
\nomenclature[V]{$x_s$}{binary decision variable for SC investment}
\nomenclature[V]{$S_s$}{capacity of SC}
\nomenclature[V]{$d_c$}{reactive current droop gain of IBR $c$ based on $S_{\mathrm{base}}$}
\nomenclature[V]{$S_{c_l}$}{capacity of GFL}
\nomenclature[V]{$S_{c_m}$}{capacity of GFM converter}
\nomenclature[V]{$S_{c_m}'$}{temporary overloading capacity of GFM converter}
\nomenclature[V]{$x_{c_m}$}{binary decision for GFM IBR investment}
\nomenclature[V]{$\lambda_{\mathrm{min}} (\cdot)$}{minimum eigenvalue}
\nomenclature[V]{$\mathbf{Y}_{eq}$}{equivalent admittance matrix}
\nomenclature[V]{$\mathbf{Y}_{red}$}{reduced node admittance matrix}
\nomenclature[V]{$P_{c_l}$}{output power of GFL IBR $c$}
\nomenclature[V]{$C_{t,n,g}$}{operation cost of SG}
\nomenclature[V]{$y_{t,n,g}^{sg}$}{start-up decision of SG}
\nomenclature[V]{$y_{t,n,g}$}{online status of SG}
\nomenclature[V]{$p_{t,n,g}$}{active power production of SG $g$}
\nomenclature[V]{$p_{t,n,w}$}{active power production of wind $w$}
\nomenclature[V]{$p_{t,n,m}$}{active power production of PV $m$}
\nomenclature[V]{$p_{t,n,c_m}$}{active power production of GFM IBR $c$}
\nomenclature[V]{$p_{t,n,l}^{c}$}{active power shedding of load $l$}
\nomenclature[V]{$\Delta Y$}{additional $Y$ matrix increment due to SGs/SCs' reactance}
\nomenclature[V]{$\widehat{I}_F$}{linearized SCC}
\nomenclature[V]{$\widehat{\mathrm{gSCR}}$}{linearized gSCR}
\nomenclature[V]{$v_{c_m}$}{binary variable for BESS charging or discharging}
\nomenclature[V]{$\mathrm{SoC}_{t,s,c_m}$}{BESS state of charge}

\printnomenclature

\section{Introduction}
Wide deployment of Inverter-Based Resources (IBRs) has been witnessed in the past few decades, to achieve decarbonization targets \cite{UK_goal} and energy independence across the world. Clean and sustainable as renewable energy is, it renders stability and security issues in power systems, because of its power electronic interface with the grid. Undesired events on system and area levels have been reported due to the decline of system inertia, grid strength, and voltage support, which are conventionally provided by Synchronous Generators (SGs). To further increase the IBR penetration level in the future grids while ensuring stable and secure system operation, the future grids must be carefully planned, given the stability issues due to the high IBR penetration. Furthermore, both the small-signal and transient stabilities can be unstable in a high IBR-penetrated system, which needs to be considered simultaneously.

Interfaced with the grid through Phase-Locked Loop (PLL), Gird-FoLlowing (GFL) IBRs suffer from small-signal stability issues in weak grids \cite{tu2021stability}. Approaches have been proposed to assess and improve the GFL stability. Reference \cite{he2021pll} proposes a simple method to efficiently assess the PLL synchronization stability level of a multi-converter system considering the interactions among different converters. A constant-coupling-effect-based PLL is proposed in \cite{lin2022constant} to improve the synchronization stability of grid-connected converters under weak grids, where the PLL bandwidth is decoupled with the converter impedance response. The authors in \cite{hu2022impedance} improve the stability of the PLL-interfaced DFIG by emulating the reference calculation matrix in the synchronous reference frame, in order to improve the phase margin while retaining the advantages of PLL under unbalanced or harmonically distorted voltage.

Although novel control strategies are proposed on the IBR device level, {sufficient system strength for small-signal stability} maintenance should also be ensured, especially at the Point of Common Coupling (PCC) of GFL IBR, at the planning stage by optimally coordinating different resources in the system. A high-level overview of the system strength from the perspective of system planning in the National Electricity Market (NEM) grid of Australia is provided in \cite{yu2022overview}, which covers the definition, attributes, and manifestations, as well as industry commentary of system strength. Various approaches and elements have been proposed and identified to enhance the system small-signal stability during the planning stage.

The authors in \cite{hadavi2021optimal} propose an optimal allocation and sizing method to determine the sizing and allocation of Synchronous Condensers (SCs) to maintain a certain system strength at points of connection. However, the optimization is solved by meta-heuristic approaches, whose global optimality cannot be guaranteed. This issue is further addressed by the same authors in \cite{hadavi2022planning} by formulating the optimization model as a mixed-integer convex problem. \textcolor{black}{An optimization algorithm is presented in \cite{de2024identifying} to meet a minimum grid strength requirement that enables the high penetration of IBRs in a weak network. The allocation of SCs to enhance the system strength is studied within the context of the Nigerian power grid  \cite{sanni2024strengthening} and National Electricity System of Chile\cite{de2023sizing}.} The work in \cite{yang2020placing} proposes an optimal allocation problem of the grid-forming converters to increase the system strength and ensure the PLL-induced small-signal stability. 

On the other hand, during the transient process, with limited fault levels \textcolor{black}{(SCC)} in the system due to the retirement of SGs, system protection, and transient stability are also compromised. Fault levels at critical nodes should always be maintained above a level so that the protection devices can be triggered properly and the voltage drops across the network during a fault do not cause the extensive trip of any other major electrical components such as SGs. 
An optimization model to retrofit the outdated thermal units into SCs is proposed in \cite{an2022enhance} to improve transient stability, which is solved through iteration. The authors in \cite{li2022voltage} present a nonlinear optimization model to determine the optimal capacity of synchronous condensers for transient voltage stability improvement. An optimization model for efficient allocation of BESS in power systems with high IBR penetration is presented in \cite{cifuentes2019network} to maximize the average voltage during faults by increasing the system strength for a set of critical scenarios. The role of Battery Energy Storage Systems (BESSs) in mitigating the voltage and frequency stability issues are investigated in \cite{ramos2023placement}, where the problem of optimal placement and sizing of the BESSs is formulated as a constrained multi-objective optimization problem, solved by binary grey wolf optimization. Reference \cite{yang2022sizing} investigates the optimal capacity and location of BESSs in a distribution to increase the stability and reliability of power systems measured by the short-term voltage stability index. Moreover, although it is generally agreed that the contribution of IBRs on the fault current is much less than that of SGs, new technologies have also been proposed to increase the temporary overcurrent capability \cite{shao2020power,ren2021phase}, which can also make the fault current provision from IBRs non-negligible.

Based on the above discussion, it is clear that different resources in the system including SGs, SCs, BESSs and IBR with Grid-ForMing (GFM) control play a crucial role in maintaining the small-signal stability measured by system strength and the fault level during the transient process. Although it has been demonstrated that synchronous units and IBRs can have similar contributions to system strength during normal operation and small disturbances, their fault behaviors are different \cite{lepour2021performance}. \textcolor{black}{Additionally, most of the existing works focus on the planning of either SCs or energy storage systems while considering a single criterion, such as system strength-based small-signal performance \cite{hadavi2022planning,richard2020optimal,hadavi2021optimal,yang2020placing} or fault level-based transient performance \cite{ramos2023placement,zhang2022short,teimourzadeh2015milp}, which may not be sufficient to maintain the overall system stability. Instead, these two aspects should be considered simultaneously, such that both the small-signal and transient performance can be enhanced.} Furthermore, even though synchronous units seem to have better performance in supporting the system stability than IBRs, the optimal solution becomes unclear considering the power/energy shifting capability of the battery storage system \cite{kalair2021role} and their enhanced overloading capability enabled by the new technologies, which requires a detailed techno-economic analysis. Moreover, most of the planning models only minimize the total investment cost but the detailed operational cost and constraints are neglected, which may lead to suboptimal or infeasible investment schemes due to a lack of operational flexibility \cite{du2018high}. 

In this context, developing a coordinated planning model that co-optimizes the sizes and locations of different resources in the system {to enhance both transient and small-signal performance} in high IBR-penetrated systems is thus necessary and beneficial. Note that the frequency stability issues due to the low inertia characteristics in high IBR-penetrated systems can be resolved by IBRs with additional control loops, which has already been extensively investigated in the literature, thus not being considered in this work. The key contributions of this work are identified below.
\begin{itemize}
    \item To enhance the transient and small-signal stability, the SCC and system strength constraints are analytically derived considering the different characteristics of synchronous units and IBRs. To evaluate the SCC while accounting for the dependence between IBR SCC injections and their terminal voltages, an iterative algorithm is further proposed with guaranteed convergence.
    \item The highly nonlinear SCC and system strength constraints are effectively linearized through a novel data-driven approach where an active sampling method is proposed to generate a representative data set considering the annual operational details, which significantly improves the linearization performance. 
    \item  A coordinated system planning model including operational details is formulated as a Mixed Integer Linear Programming (MILP), to improve both small-signal and transient stability in weak grids, by co-optimizing the sizes and locations of SCs and GFM IBRs with overloading capabilities.  
    \item The effectiveness and scalability of the proposed coordinated planning model are demonstrated through case studies based on IEEE 39- and 118-bus systems, with significant cost savings and the impact of the IBR's overloading capability illustrated.
\end{itemize}
The remaining part of this paper is organized as follows. Section~\ref{sec:2} models the synchronous units and IBR for SCC calculation. Section~\ref{sec:3} introduces the system strength metric considered in this work, followed by system planning model formulation in Section~\ref{sec:4}. Case studies are presented in Section~\ref{sec:5}. Finally, Section~\ref{sec:6} concludes the paper.

\section{Short Circuit Current Quantification} \label{sec:2}
To enhance the transient performance, sufficient SCC should be maintained at critical locations to ensure that the protection devices can be triggered properly and the voltage drops across the network during a fault do not cause the extensive trip of any other major electrical components such as SGs.  To achieve this, the SCC is analytically derived as an explicit function of the decision variables in system operation and planning stages, while considering the different characteristics of synchronous units and IBRs. Utilizing the SCC-based indices to characterize the transient stability is also common in literature, e.g., \cite{ghamsari2021linearized,machowski2015simplified}.

\subsection{Modeling of Synchronous Units}
The conventional model of SGs and SCs for short current calculation is the same. Therefore, only the example of SGs is given here. An SG $g\in\mathcal{G}$ is modeled as a voltage source $E_g$ behind a reactance $X_{g}$, which can be further converted to a current source $I_g$ in parallel with a susceptance $Y_g$ according to the Norton's theorem:
\begin{align}
    I_g & = \frac{E_g }{X_{g} } \label{Ig} \\
    \label{Yg}
    Y_g & = \frac{1}{\mathrm{j} X_{g} }.
\end{align}
Replacing the SG index $g\in\mathcal{G}$ in \eqref{Yg} with that of SCs $s\in\mathcal{S}$ gives the expression of the SC model. It can be observed that synchronous units influence the SCC by reshaping the system admittance matrix through $Y_{g}$ and $Y_s$.

\subsection{Modeling of IBRs}
The saturation of IBR current output indicates that the model of a fixed voltage behind a reactance like an SG is inappropriate for an IBR. Instead, an IBR can be modeled as a voltage-dependent current source that injects SCC depending on its terminal voltage according to the implemented droop control.

Combining the classic SCC superposition approach with the IBR model enables the SCC calculation in a general power system with both SGs and IBRs. The SCC at a fault bus $F$ ($I_F$) can be computed through KCL in the pure-fault system where the only sources are those at the IBR buses and the fault bus \cite{9329077}:
\begin{equation}
\label{-V_F(0)}
    -V_F(0) = \sum_{c\in \mathcal{C}} Z_{F\Phi(c)}(I_{fc}-I_{Lc})+Z_{FF}I_{F},
\end{equation}
where $V_F(0)$ is the pre-fault voltage at Bus $F$; $Z$ is the system impedance matrix, which also includes the impedance from SGs and SCs as defined in Section~\ref{sec:4.2}; $\Phi(c)$ maps the IBR $c\in \mathcal{C}$ to the corresponding bus index; $I_{fc}$ and $I_{Lc}$ are the fault current and pre-fault load current from IBR $c$ respectively. Rearranging \eqref{-V_F(0)} yields the expression of the SCC at Bus $F$ as follows:
\begin{equation}
\label{I_sc1}
    I_{F}  = \frac{-V_F(0)-\sum_{c\in \mathcal{C}} Z_{F\Phi(c)}(I_{fc}-I_{Lc})}{Z_{FF}}.
\end{equation}
The fault current from IBR $c$ is injected according to its terminal voltage drop, which can be modeled as a voltage-dependent current source. The UK national grid requires a reactive current of full capacity (1.0 - 1.5 $\mathrm{p.u.}$) from all IBRs when their terminal voltages drop to zero \cite{NG_GridCode}. Hence, the fault current from IBR $c\in \mathcal{C}$ can be calculated as:
\begin{equation}
\label{I_fc_droop}
    I_{fc} = -\textsf{j} d_c\big({|V_{\Phi(c)}|-|V_{\Phi(c)}(0)|}\big),
\end{equation}
where $d_c$ is the reactive current droop gain; $V_{\Phi(c)}$ and $V_{\Phi(c)}(0)$ are the post-fault and pre-fault voltage at bus $\Phi(c)$. Based on the superposition principle, the voltage drop at bus $\Phi(c)$, $\Delta V_{\Phi(c)}$ can be derived:
\begin{equation}
\label{Delta_V}
    \Delta V_{\Phi(c)} = \sum_{c'\in \mathcal{C}} Z_{\Phi(c')\Phi(c)}(I_{fc'}-I_{Lc'})+Z_{F\Phi(c)}I_{F}.
\end{equation}
Equation \eqref{Delta_V} is essentially an implicit function of $\Delta V_{\Phi(c)}$, due to the dependence of $I_{fc}$ on $\Delta V_{\Phi(c)}$. By combining \eqref{I_fc_droop} and \eqref{Delta_V}, $\forall c\in\mathcal{C}$ and neglecting the pre-fault load current and line resistance, the following expression of $\Delta V_{\mathcal{C}}$ can be obtained:
\begin{equation}
\label{Delta_V_explicit}
    \Delta V_{\mathcal{C}} = A_Z^{-1}
    Z_{F\mathcal{C}}I_{F},
\end{equation}
where $\Delta V_{\mathcal{C}} \in \mathbb R^{|\mathcal{C}|}$ is the vector of IBR terminal voltage deviations; $ Z_{F\mathcal{C}} \in \mathbb R^{|\mathcal{C}|}$ is the vector collecting $Z_{F\Phi(c)},\,\forall c\in\mathcal{C}$ in the $Z$ matrix; $A_Z\in \mathbb R^{|\mathcal{C}|\times|\mathcal{C}|}$ is defined as:
\begin{equation}
\label{A_Z}
    A_{Z_{ij}}=
    \begin{cases}
    Z_{\Phi(i)\Phi(i)}\textsf{j}d_i+1\;\;&\mathrm{,if}\,i = j \\
    Z_{\Phi(j)\Phi(i)}\textsf{j}d_j\;\;&\mathrm{,if}\,i \neq j
    \end{cases}.
\end{equation}

Combining \eqref{I_sc1}, \eqref{I_fc_droop} and \eqref{Delta_V_explicit} gives the SCC at fault bus $F$:
\begin{equation}
\label{I_sc2}
    I_{F} = \frac{-V_F(0)+\sum_{c\in\mathcal{C}}{Z_{F\Phi(c)}I_{Lc}}}{Z_{FF}-\textsf{j}Z^{\mathsf T}_{F\mathcal{C}}\mathrm{diag}(d_c)A_Z^{-1}
    Z_{F\mathcal{C}}},
\end{equation}
with $\mathrm{diag}(d_c)\in \mathbb R^{|\mathcal{C}|\times|\mathcal{C}|}$ being the diagonal matrix with $d_c$ being the diagonals. After neglecting the pre-fault load current, the SCC constraint can be expressed as:
\begin{equation}
\label{I_sc3}
    \left|I_{F} \right|= \frac{\left|V_F(0)\right|}{\left|Z_{FF}-\textsf{j}Z^{\mathsf T}_{F\mathcal{C}}\mathrm{diag}(d_c)A_Z^{-1}
    Z_{F\mathcal{C}}\right|} \ge I_{F_{\mathrm{lim}}}.
\end{equation}
Note that the above SCC expression reduces to the conventional formula, $V_F(0)/Z_{FF}$, if the SCC from IBR units is neglected, i.e., $d_c=0$.


However, there are grid codes in other countries with different requirements regarding the fault current contribution from IBRs, such as full capacity reactive current at $0.5\,\mathrm{p.u.}$ terminal voltage drop \cite{ALSHETWI2020119831}. The fault current from IBRs as previously defined in \eqref{I_fc_droop} is then modified as:
\begin{equation}
\label{I_fc_droop_1}
    |I_{fc}| = \min\left\{I_c^{\mathrm{max}}, d_c\Delta \left|V_{\Phi(c)} \right| \right\}.
\end{equation}
In this case, a simple explicit expression of the SCC as in \eqref{I_sc2} may not be available due to the interdependence between the IBR fault currents and post-fault voltages. Nevertheless, $I_{F}$ can still be calculated iteratively, given the system operating conditions as demonstrated in Algorithm~\ref{alg:SCC}. The key idea is to initialize the IBR terminal voltages to $-1$ (Step 2) and update the IBR current (Step 5) and then the voltage (Step 7) with the latest updated voltage drop and fault current until the error of two successive steps is smaller than a predefined limit (Step 9). The convergence of the algorithm is proved as follows.
\color{black}
\begin{proof}
    First, the bound of the sequence can be directly obtained by its definition: $\Delta |V^{(k)}_{\Phi(c)}| \in [-1,0]$, since we assume that the pre-fault voltage in transmission level is approximate $1 \mathrm{p.u.}$. 
    Second, we prove the sequence $\Delta |V^{(k)}_{\Phi(c)}|$ is monotonically increasing by induction:
    For $k = 0$ the statement, $\Delta |V^{(0)}_{\Phi(c)}|\le\Delta |V^{(1)}_{\Phi(c)}|$ is true, since $\Delta |V^{(0)}_{\Phi(c)}|=-1$ and $\Delta |V^{(k)}_{\Phi(c)}| \in [-1,0]$. Further suppose $\Delta |V^{(k)}_{\Phi(c)}|\le\Delta |V^{(k+1)}_{\Phi(c)}|$ for some $k\in\mathbb N$. Then,

\begin{align*}
    \Delta |V_{\Phi(c)}^{(k+2)}| & = \sum_{c'\in \mathcal{C}} Z_{\Phi(c')\Phi(c)}(I_{fc'}^{(k)}-I_{Lc'})+Z_{F\Phi(c)}I_{F}^{(k+1)} \\
    & \ge \sum_{c'\in \mathcal{C}} Z_{\Phi(c')\Phi(c)}(I_{fc'}^{(k-1)}-I_{Lc'})+Z_{F\Phi(c)}I_{F}^{(k)} \\
    & = \Delta |V_{\Phi(c)}^{(k+1)}|,
\end{align*}
    where the inequality can be proved by combining (4), (11) and $\Delta |V^{(k)}_{\Phi(c)}|\le\Delta |V^{(k+1)}_{\Phi(c)}|$. The convergence of $\Delta |V_{\Phi(c)}^{(k)}|$ is proved by the monotone convergence theorem, thus concluding the proof.
    \end{proof}
\color{black}


\begin{algorithm}[!b]
\caption{Iterative calculation of SCC}
\label{alg:SCC}
\begin{algorithmic}[1]
    \State Set $k=0$ and $\varepsilon=0$
    \State Initialization
    \Comment $\Delta |V^{(0)}_{\Phi(c)}| = -1$
    \While {$\varepsilon>\epsilon$ or $k=0$}
        \State $k = k + 1$
        \State Calculate $I_{fc}^{(k-1)}=\min\left\{I_c^{\mathrm{max}}, d_c \Delta \left|V_{\Phi(c)}^{(k-1)} \right| \right\},\,\forall c$ 
        \State Calculate $I_F^{(k)}=f\left(I_{fc}^{(k-1)}\right)$,  according to \eqref{I_sc1}
        \State Calculate $\Delta V^{(k)}_{\Phi(c)}=g\left(I_{fc}^{(k-1)},I_F^{(k)}\right)$,  according to \eqref{Delta_V}
                
        \State Compute error terms
        \Comment $\varepsilon_I=\left|I_F^{(k)}-I_F^{(k-1)}\right|$
        
        \Comment $\varepsilon_V=\left|\Delta |V^{(k)}_{\Phi(c)}|-\Delta |V^{(k-1)}_{\Phi(c)}|\right|$
        \State Determine convergence error 
        \Comment $\varepsilon=\varepsilon_I+\varepsilon_V$
	\EndWhile
	\State Return $I_F^{(k)}$
\end{algorithmic}
\end{algorithm}

\subsection{Incorporating decision variables of the planning model}\label{sec:2.3}
In order to embed the SCC constraint derived in \eqref{I_sc3} into the system planning problem, it is necessary to demonstrate the connection between \eqref{I_sc3} and the decision variables. Furthermore, all quantities associated with generating units defined in previous sections are per unit values based on the capacity of their own, which also need to be converted to a global base, $S_{\mathrm{base}}$. For conventional SGs, their online statuses are viewed as decision variables, to account for the operational details in the planning model. Hence, the SCC model defined in \eqref{Ig} and \eqref{Yg} are modified as follows:
\begin{align}
    I_g & = \frac{E_g S_g}{X_{g} S_{\mathrm{base}}} x_g \\
    Y_g & = \frac{S_g}{\mathrm{j} X_{g} S_{\mathrm{base}}} x_g,
\end{align}
where $S_g$ is the capacity of synchronous generator $g\in \mathcal{G}$ and $x_g\in \{0,1\}$ is the binary decision variable, representing its online state. Similarly, for SC $s\in \mathcal{S}$, both the capacity and location are decision variables in the planning model, which can be expressed as below:
\begin{align}
    I_s & = \frac{E_s S_s}{X_{s} S_{\mathrm{base}}} \\
    Y_s & = \frac{S_s}{\mathrm{j} X_{s} S_{\mathrm{base}}}.
\end{align}
$S_s$ is the decision variable related to the SC's capacity and location:
\begin{subequations}
\label{S_s}
\begin{align}
    \underline{S}_s x_s \leq & S_s\leq \Bar{S}_s x_s\\
    \sum_{s\in\mathcal{S}} x_s\leq & \Bar{N}_s,
\end{align}
\end{subequations}
with $\underline{S}_s$, $\Bar{S}_s$ being the lower and upper capacity limits of the potential SC $s\in \mathcal{S}$ and $x_s\in \{0,1\}$ the binary decision of whether the SC should be installed. $\Bar{N}_s$ is the maximum number of SCs that can be installed. \textcolor{black}{Here, we assume that the SC capacity can be selected continuously from $\underline{S}_s$ and $\Bar{S}_s$. However, in reality, if only discrete sizes could be selected as indicated by the reviewer, the lowest possible value of the SC capacity above the optimal solution should be selected to ensure conservativeness. In this case, the larger the discrete size, the less conservative the results are.}

As for the IBRs, the online capacities of the GFL IBRs, $c \in \mathcal{C}_l$ are decision variables during the operation stage whereas the investment capacities of the BESSs (GFM IBRs, $c \in \mathcal{C}_m$) are decision variables during the planning stage. Moreover, to account for the investment of the IBR temporary overloading capability, the decisions on the BESS capacities ($S_{c_m}$) and its temporary overloading capacities ($S_{c_m}'$) are treated differently. As a result, the IBR droop control gain $d_c,\,\forall c \in \mathcal{C}$ can be re-scaled as follows:
\begin{equation}
\label{scale_droop}
    d_c = \begin{cases}
    d_{0}\times\frac{S_{c_l}}{S_{\mathrm{base}}}\;\;&\mathrm{if}\,c\in\mathcal{C}_l\\
    d_{0}\times\frac{S_{c_m}'}{S_{\mathrm{base}}}\;\;&\mathrm{if}\,c\in\mathcal{C}_m,
    \end{cases}
\end{equation}
where $d_{0}$ is the droop gain based on the IBR's own capacity and $S_{c_l}$, $S_{c_m}'$ are the capacity of GFL, the temporary overloading capacity of GFM converter respectively. For GFL units, $c\in\mathcal{C}_l$, their locations, and installed capacities are fixed depending on the renewable resources, whereas for the GFM units, $c\in\mathcal{C}_m$, their locations and capacities are decision variables at the system planning stage. The temporary overloading capacity $S_{c_m}'$ can be further expressed as:
\begin{equation}
    S_{c_m} \leq S_{c_m}' \leq \beta S_{c_m},
\end{equation}
where $\beta$ is the temporary overloading factor. $S_{c_m}$ is the GFM capacity during normal operation being confined by:
\begin{subequations}
\label{S_cm}
\begin{align}
    \underline{S}_{c_m} x_{c_m} \leq & S_{c_m}\leq \Bar{S}_{c_m} x_{c_m}\\
    \sum_{c_m\in\mathcal{C}_m} x_{c_m}\leq & \Bar{N}_{c_m},
\end{align}    
\end{subequations}
where $\underline{S}_{c_m}$, $\Bar{S}_{c_m}$ is the lower and upper limits of $S_{c_m}$, $x_{c_m}\in \{0,1\}$ is the binary decision of whether the GFM IBR should be installed and $\Bar{N}_{c_m}$ is the maximum number of GFM IBR that can be installed. Thus far, it is clear that the commitment decisions of SGs, the investment decisions of SCs, and GFM IBRs (BESSs) including both locations and capacities influence the system SCC.

\section{System Strength Assessment} \label{sec:3}
In order to ensure the small-signal stability of GFL IBRs, the system strengths at their terminal buses should be kept above a certain limit. Conventionally, system strength is assessed through Short Circuit Ratio (SCR), which is a locational property, being defined as the ratio of three-phase short circuit capacity to the IBR rated power. However, as the IBR penetration increases, the effectiveness of using SCR to represent system strength becomes questionable due to the neglect of IBR interactions. As a result, extensive research has been conducted in this area to define more appropriate system strength index for IBR stability characterizing, such as weighted short circuit ratio developed by ERCOT \cite{schmall2015voltage}, the composite short circuit ratio developed by GE \cite{fernandes2015report} and the Equivalent Short Circuit Ratio (ESCR) developed in \cite{kim2022evaluation}. In this work, the generalized Short Circuit Ratio (gSCR) in \cite{liu2023generalized} is utilized as the system strength index in the system planning model to ensure the IBR stability. The gSCR is defined as the minimum eigenvalue of the equivalent admittance matrix, $\mathbf{Y}_{eq}$:
\begin{subequations}
\label{gSCR}
\begin{align}
    \mathrm{gSCR} &= \lambda_{\mathrm{min}} (\mathbf{Y}_{eq}) \\
    \mathbf{Y}_{eq} &= \mathrm{diag}\left(\frac{V^2_{\Phi(c_l)}}{P_{c_l}}\right) \mathbf{Y}_{red} \label{Yeq},
\end{align}
\end{subequations}
where $\mathrm{diag}\left({V^2_{\Phi(c_l)}}/{P_{c_l}}\right)$ is the diagonal matrix related to the GFL IBR terminal voltage and output power and $\mathbf{Y}_{red}$ is the reduced node admittance matrix after eliminating passive buses. \textcolor{black}{The above small-signal stability index considers the impact of GFL (dynamics and locations) and the networks (also including the impedance of SGs, SCs and GFMs). The GFL dynamics influences the value of $\mathrm{gSCR}_{\mathrm{lim}}$, whereas the GFL capacity and location as well as the network influence the value of $\mathrm{gSCR}$, by influencing $\mathbf Y_{eq}$. In other words, the stability analysis based on the above gSCR-based approach decouples the dynamics-related quantities (functions of $s$) with the steady-state quantities ($\mathbf{Y}_{eq} =\mathrm{diag}\left(\frac{V^2_{\Phi(c_l)}}{P_{c_l}}\right) \mathbf{Y}_{red}$). The former influences $\mathrm{gSCR}_{\mathrm{lim}}$ and is determined by the GFL dynamics and the network dynamics, which are fixed at the system operation stage once the GFL control algorithm and parameters are selected. The latter influences $\mathrm{gSCR}$ and is determined by the GFL capacity and location as well as the admittance matrix (including SGs, SCs and GFMs). As a result, by forcing the $\mathrm{gSCR}\le \mathrm{gSCR}_{\mathrm{lim}}$ during system operation, the small signal stability can be maintained. }


Different from the impedance matrix $Z$ used for SCC calculation, which includes the contribution from SGs and SCs, the admittance matrix $\mathbf{Y}$ based on which $\mathbf{Y}_{eq}$ in \eqref{gSCR} is calculated, should also include the admittance of GFM IBRs as demonstrated in Section~\ref{sec:4.2.2}. This is because during the transient process (short circuit), both GFM and GFL IBRs are modeled as voltage-dependent current sources due to the saturation of inverter current, whereas for the system strength assessment, since all of the indices mentioned above are derived from the perspective of small signal stability, a GFM IBR can thus be modeled the same as a synchronous unit, i.e., a voltage source behind impedance due to its grid-forming capability \cite{yu2022overview,yang2020placing}. As a result, the system strength measured by gSCR can be enhanced by having more SG/SC and GFM IBR in the system or by reducing the output power from GFL units. It should be noted that the focus of this work is not the derivation of the system strength index. Instead, we utilize an established index to characterize the system strength in the system planning model. Although, the specific form in \eqref{gSCR} is chosen, the proposed method could in general deal with other system strength indices such as the composite short circuit ratio in \cite{fernandes2015report} and the equivalent effective short circuit ratio in \cite{kim2022evaluation}. 

\textcolor{black}{It can be observed that SCC and gSCR are influenced by the online capacity of SGs, which can be shaped during normal operation by system scheduling. Furthermore, in the case where additional SCs and/or GFM converters are needed to support the system stability, the system planning model is necessary. Therefore, this work intends to determine the optimal configuration (size and location) of the SCs and GFM IBRs at the system planning stage as well as the operation status of SGs during normal operation by including the operational details in the planning model, as demonstrated in the next section. By formulating system dynamic performance as operational constraints, the system transient regime can be implemented in the planning model.}

\section{System Planning Problem} \label{sec:4}
In this section, the proposed system planning model is mathematically formulated, where the investment of SCs and GFM IBRs (BESSs) as well as the temporary overloading capability of the latter are optimally determined to maintain the required SCC level and system strength. In addition, detailed operational constraints, including the SG ramps, start-up/shut-down limits, and transmission constraints are embedded into the optimization problem to model the operational characteristics of a high IBR-penetrated system. The overall planning model is formulated as an MILP with the reformulation of the SCC and system strength constraint illustrated in this section. 

\textcolor{black}{The proposed model is intended for system operators who are responsible for delivering affordable, clean and secure power during the transition toward the future net zero system. However, different from the conventional generation/transmission expansion planning problem, we only focus on the optimal placement of the SCs and GFM BESSs to enhance system stability, based on which we demonstrate the possibility and importance of including both small-signal and transient constraints in the planning model as well as their impact on the investment decisions. The system operator carries out the proposed method in a centralized manner with all the necessary information such as future demand and generation, the potential location and feasible capacity range of different system assets, such that different resources in the system including SCs and GFM IBRs with overloading capabilities can be optimally installed to improve both small-signal and transient performance of the system.}

\textcolor{black}{It should also be noted that the proposed model is not intended to replace or to be more adequate/beneficial than the conventional generation/transmission expansion planning model, the main purpose of which is to ensure that the power system can meet future electricity demand cost-effectively. Instead, this work only focuses on the optimal placement of the SCs and GFM BESSs to enhance system stability, based on which we demonstrate the importance of including both small-signal and transient constraints in the planning model and their impact on the investment decisions. In future work, we would also like to investigate how the proposed model can be combined with the generation/transmission expansion planning model.}

\subsection{Objective Function} \label{sec:3.1}
The objective of the proposed planning model is to minimize the sum of annualized capital investment costs and the expected yearly operating cost over all nodes in a given scenario tree:
\begin{subequations}
    \label{eq:SUC}
\begin{align}
    \min &\sum_{s\in \mathcal{S} }c^{s} S_{s} + \sum_{c\in \mathcal{C} }c^c S_c + \nonumber \\
    &\sum_{t\in T} \sum_{n\in \mathcal{N}} \pi_n \left( \sum_{g\in \mathcal{G}}  C_{t,n,g} + \Delta t \sum_{l\in \mathcal{L}} c^{VOLL} p_{t,n,l}^{c} \right) \\
    & \color{black} C_{t,n,g} = c_g^{st} y_{t,n,g}^{sg} + c_g^{nl} y_{t,n,g} + c_g^{m} p_{t,n,g} \label{thermal_cost},
\end{align}    
\end{subequations}
where $c^{s}$ (or $c^c$) is the annualized investment cost coefficient of SC (or GFM IBR converter). \textcolor{black}{Note that the BESS degradation cost is not explicitly considered in the model. Since the additional stability constraints have little impact on the BESS operation (charging and discharging) behavior. On one hand, the support of the BESS on the small-signal stability comes from the voltage source nature of the GFM, which does not change the charging and discharging behavior of the BESS. On the other hand, although the transient support requires the BESS to provide reactive current up to its capacity during faults, they are very rare events in power systems. Nevertheless, accurate modeling of the degradation process in the BESS planning problem is out of the scope of this work and can be incorporated into the proposed model by utilizing the methods presented in the existing research such as \cite{amini2021optimal}.} $\pi_n$ is the probability of scenario $n\in \mathcal{N}$ and \textcolor{black}{$C_{t,n,g}$ is the operation cost of unit $g\in \mathcal{G}$ in scenario $n\in\mathcal{N}$ at time step $t\in T$ including startup, no-load and marginal cost as indicated in \eqref{thermal_cost} with $c_g^{st}$, $c_g^{nl}$, $c_g^{m}$ being the cost coefficients and $y_{t,n,g}^{sg}$, $y_{t,n,g}$, $p_{t,n,g}$ being the start-up decision, online status, and active power production respectively}; $c^{VOLL}$ represents the value of lost load; $p_{t,n,l}^{c}$ is the active power shedding of load $l$ in scenario $n$ at time step $t$. \textcolor{black}{The scenario tree is built based on user-defined quantiles of the forecasting error distribution to capture the uncertainty associated with wind generation. Two steps are involved to generate the scenarios: (i) creating the distribution of net demand;  (ii) quantifying the value of net demand and its probability in each node. The readers are referred to \cite{7115982} for more information. Also note that although there has been more recent research to manage the uncertainty at the system planning stage, since the uncertainty management is not the focus of this work, these methods are not applied or discussed in the paper.}

\subsection{Short Circuit Current and System Strength Constraints}\label{sec:4.2}
In this section, the short circuit current and system strength expressions previously explained in Section~\ref{sec:2} and Section~\ref{sec:3} are reformulated as linear constraints, which can be directly included in the system planning model. \textcolor{black}{We assume that the disturbance could potentially happen at any time during normal operation, and hence the dynamic performance is formulated as operational constraints in the planning model. In this way, by adjusting the normal operating conditions as well as the investment decisions, the system dynamic performance can be improved in a cost-effective manner.}

\subsubsection{Short circuit current constraints}
Combining \eqref{I_sc3} and \eqref{scale_droop} gives the SCC constraint: 
\begin{equation}
\label{I_sc4}
    \left|I_{F}  \right|= \frac{\left|V_F(0)\right|}{\left|Z_{FF}-\textsf{j}Z^{\mathsf T}_{F\mathcal{C}}\mathrm{diag}(d_0S_c/S_{\mathrm{base}})A_Z^{-1}
    Z_{F\mathcal{C}}\right|} \ge I_{F_{\mathrm{lim}}},
\end{equation}
where $S_c = S_{c_l},\,\forall c\in \mathcal{C}_l$ and $S_c = S_{c_m}',\,\forall c\in \mathcal{C}_m$. The SCC is directly influenced by the elements in the impedance matrix $Z$ and the converter capacity $S_c$. The $Z$ matrix, by definition, can be obtained by taking the inverse of the system admittance matrix $Y$ as follows:
\begin{subequations}
\label{Y}
    \begin{align}
    Z &= Y^{-1}\\
    Y &= Y^0 +  \Delta Y,
    \end{align}
\end{subequations}
where $Y^0$ is the admittance matrix of the transmission lines only; $\Delta Y$ denotes the additional $Y$ matrix increment due to SGs/SCs' reactance. Depending on the operating conditions of the SGs and the investment decision of SCs, the elements in $\Delta Y$ can be expressed as:
\begin{equation}
\label{Y2}
    \Delta Y_{ij}=
    \begin{cases}
    \frac{S_g}{X_{g}S_{\mathrm{base}}}x_{g}\;\;&\mathrm{if}\,i = j \land \exists\, g\in \mathcal{G},\, \mathrm{s.t.}\,i=\Psi(g)\\
    \frac{S_s}{X_{s}S_{\mathrm{base}}}\;\;&\mathrm{if}\,i = j \land \exists\, s\in \mathcal{S},\, \mathrm{s.t.}\,i=\Psi(s)\\
    0\;\;& \mathrm{otherwise},
    \end{cases}
\end{equation}
where $S_s$ is confined by \eqref{S_s} and $\Psi(\cdot)$ maps the synchronous unit $g\in \mathcal{G}/s\in \mathcal{S}$ to the corresponding bus index. It should be noted that $x_{g},\,\forall g \in \mathcal{G}$ can be viewed as binary decision variables, whereas the SC capacity ($S_s$) in \eqref{Y2} and IBR capacity ($S_c$) in \eqref{I_sc4} are deemed as continuous variables, which makes \eqref{I_sc4} a highly nonlinear constraint involving matrix inverse with binary and continuous decision variables. 

\subsubsection{System strength constraints} \label{sec:4.2.2}
Similarly, system strength constraint at GFL IBR $c_l$ can be written as:
\begin{equation}
\label{ESCR_lim}
    \mathrm{gSCR} =\lambda_{\mathrm{min}} \left(\mathrm{diag}\left(\frac{V^2_{\Phi(c_l)}}{P_{c_l}}\right) \mathbf{Y}_{red} \right) \ge \mathrm{gSCR}_{\mathrm{lim}},
\end{equation}
with $\mathrm{gSCR}_{\mathrm{lim}}$ being the gSCR limit that ensures the IBR's stability. Similarly, the admittance matrix for system strength evaluation before node-reduction, $\mathbf{Y}$ can be derived:
\begin{subequations}
\label{mathbf_Y}
    \begin{align}
    \mathbf{Y} = \mathbf{Y}^0 +  \Delta \mathbf{Y}.
    \end{align}
\end{subequations}
During normal operation and small disturbances SGs, SCs, and GFM IBRs can be viewed as voltage sources behind impedances. Therefore, depending on the operating conditions of the SGs and the investment decision of SCs and GFM IBRs, the elements in $\Delta \mathbf{Y}$ can be expressed as:
\begin{equation}
\label{mathbf_Y2}
    \Delta  \mathbf{Y}_{ij}=
    \begin{cases}
    \frac{S_g}{X_{g}S_{\mathrm{base}}}x_{g}\;\;&\mathrm{if}\,i = j \land \exists\, g\in \mathcal{G},\, \mathrm{s.t.}\,i=\Psi(g)\\
    \frac{S_s}{X_{s}S_{\mathrm{base}}}\;\;&\mathrm{if}\,i = j \land \exists\, s\in \mathcal{S},\, \mathrm{s.t.}\,i=\Psi(s)\\
    \frac{S_{c_m}}{X_{c_m}S_{\mathrm{base}}}\;\;&\mathrm{if}\,i = j \land \exists\, {c_m}\in \mathcal{C}_m,\, \mathrm{s.t.}\,i=\Phi(s)\\
    0\;\;& \mathrm{otherwise},
    \end{cases}
\end{equation}
where $X_{c_m}$ is the GFM reactance; $S_s$, $S_{c_m}$ are the planning decisions associated with SC and GFM IBR capacity, constrained by \eqref{S_s} and \eqref{S_cm} respectively.

\subsubsection{Constraint reformulation} \label{sec:4.2.3}
As demonstrated in previous sections, the SCC and system strength constraints \eqref{I_sc4}-\eqref{mathbf_Y2} are highly nonlinear and involve decision-dependent matrix inverse. Although it is possible to theoretically derive each element in the system impedance matrix as a function of the decision variables, the expression becomes extremely complicated in a general multi-bus system, thus being problematic to be directly included in the system planning problem. To effectively linearize the SCC and system strength constraints, the novel data-driven approach proposed in \cite{9329077} is adapted. First, define the linearized expression of \eqref{I_sc4} and \eqref{ESCR_lim}:
\begin{align}
\label{SCC_linear}
    \widehat{I}_F & = \sum_{g\in \mathcal{G}} k_g' x_g + \sum_{s\in \mathcal{S}} k_s' S_s + \sum_{c_m\in \mathcal{C}_m} k_{c_m}' S_{c_m} \nonumber \\
    & + \sum_{c_l\in \mathcal{C}_l} k_{c_l}' P_{c_l} + k_0' \ge I_{F_\mathrm{lim}},
\end{align}
\begin{align}
\label{ESCR_linear}
    \widehat{\mathrm{gSCR}} & = \sum_{g\in \mathcal{G}} k_g x_g + \sum_{s\in \mathcal{S}} k_s S_s + \sum_{c_m\in \mathcal{C}_m} k_{c_m} S_{c_m} \nonumber\\
    & + \sum_{c_l\in \mathcal{C}_l} k_{c_l} P_{c_l} + k_0 \ge \mathrm{gSCR}_{\mathrm{lim}},
\end{align}
with $\widehat{I}_F$/$\widehat{\mathrm{gSCR}}$ being the linearized SCC/gSCR and $\mathcal{K}' =\{k_g', k_s', k_{c_m}', k_{c_l}', k_0'\}$/$\mathcal{K} =\{k_g, k_s, k_{c_m}, k_{c_l}, k_0\}$ the associated linear coefficient. Due to the similarity of SCC and system strength constraints, only an example of the latter is presented here. In order to determine the optimal parameters $\mathcal{K}$, the following optimization is solved, $\forall c_l \in \mathcal{C}_l$:
\begin{subequations}
\label{DM3}
\begin{align}
    \label{obj3}
    \min_{\mathcal{K}}\quad & \sum_{\omega \in \Omega_2} \left({\mathrm{gSCR}}^{(\omega)} - \widehat{\mathrm{gSCR}}^{(\omega)} \right)^2\\
    \label{coef_ctr2}
    \mathrm{s.t.}\quad & \widehat{\mathrm{gSCR}}^{(\omega)}< \mathrm{gSCR}_{\mathrm{lim}},\,\,\forall \omega \in \Omega_1\\
    \label{coef_ctr3}
    &\widehat{\mathrm{gSCR}}^{(\omega)} \ge \mathrm{gSCR}_{\mathrm{lim}},\,\,\forall \omega \in \Omega_3,
\end{align}
\end{subequations}
with $(\cdot)^{(\omega)}$ denoting quantities associated with sample $\omega$, and $\omega = \{x_g^{(\omega)},S_s^{(\omega)},S_{c_m}^{(\omega)}, P_{c_l}^{(\omega)}, {\mathrm{gSCR}}^{(\omega)}\}\in \Omega$ denoting the entire data set corresponding to the gSCR constraint. It is generated by evaluating $\mathrm{gSCR}$ in representative system conditions. To obtain a representative data set with a finite size, a novel active sampling method is proposed in Section~\ref{sec:4.2.4}.

The sets $\Omega_1 ,\, \Omega_2$ and $\Omega_3 $ are the subsets of $\Omega$, whose relationship is defined as below:
\begin{subequations}\label{Omega}
\begin{align}
    \Omega &= \Omega_1 \cup\Omega_2\cup\Omega_3 \\
    \label{Omega1}
    \Omega_1 & = \left\{\omega\in \Omega \mid \mathrm{gSCR}<\mathrm{gSCR}_{\mathrm{lim}} \right\}\\
    \label{Omega2}
    \Omega_2 & = \left\{\omega\in \Omega \mid \mathrm{gSCR}_{\mathrm{lim}} \le \mathrm{gSCR}<\mathrm{gSCR}_{\mathrm{lim}} + \nu \right\}\\
    \label{Omega3}
    \Omega_3 & = \left\{\omega\in \Omega \mid \mathrm{gSCR}_{\mathrm{lim}} + \nu\le \mathrm{gSCR} \right\},
\end{align}
\end{subequations}
with $\nu$ being a constant parameter. Given \eqref{coef_ctr2} and \eqref{Omega1}, all the data points whose real gSCR is smaller than the limit can be identified correctly by the estimated function, $\widehat{\mathrm{gSCR}}$. Ideally, it is also desired to correctly identify all the above-limit data points, which would make the problem become a classification model. However, this may cause infeasibility due to the restricted linear structure defined in \eqref{ESCR_linear}. Therefore, a parameter $\nu\in \mathbb{R}^+$ is introduced to define $\Omega_2$ and $\Omega_3$ as in \eqref{Omega2} and \eqref{Omega3}. In this way, all the data points in $\Omega_3$ will be classified correctly and misclassification can only occur in $\Omega_2$, thus being conservative. Furthermore, $\nu$ should be chosen as small as possible while ensuring the feasibility of \eqref{DM3}. 

\subsubsection{Active Sampling} \label{sec:4.2.4}
\textcolor{black}{In the original approach proposed in \cite{9329077}, since only binary decision variables are considered, the data set is generated by evaluating the performance index at all possible combinations. However, the stability index $\mathrm{gSCR}$ not only depends on the generator status but also some other operating conditions such as the active power injections. As a result, both binary and continuous variables are of concern and a finite data set cannot be generated by the original approach.} To obtain a representative data set with a finite size, based on which the coefficients $\mathcal{K}$ can be optimally determined from \eqref{DM3}, Algorithm~\ref{alg:sample} is proposed, which is summarized as follows. Initialize $\mathcal{K}^0$ such that the constraint \eqref{ESCR_linear} becomes redundant in the first iteration ($m = 0$) and all the data points during the entire time horizon obtained by solving the planning problem $\mathcal{P}$ are included in the set $\Omega^0$ (Step~7\,\&\,8). In each of the following iteration ($1\le m\le m^{\mathrm{max}}$), solve the planning problem with the latest updated coefficients ($\mathcal{K}^m$) and increase the data set by $\Delta \Omega$, which includes all the data points that do not satisfy \eqref{coef_ctr2} and \eqref{coef_ctr3} (misclassified), as in Step~4\,-\,6\,\&\,10. The process terminates until $\Delta \Omega = \text{\O}$ or $m = m^{\mathrm{max}}$.

\begin{algorithm}[!t]
\caption{Active Sampling}
\label{alg:sample}
\begin{algorithmic}[1]
    \State Set $m=0$, $\Delta \Omega = \{1\}$ and $k=0, \forall k \in \mathcal{K}^0$
    \State Initialization
    \Comment $\mathcal{K}^0 = \{0,0,0,0,\mathrm{gSCR}_{\mathrm{lim}}\}$
    \While {$m\le m^{\mathrm{max}}$ $\land$ $|\Delta \Omega|>0$}
        \State Solve $\mathcal{P}$ with $\mathcal{K}^m$, over the entire horizon $t\in \mathcal{T}$,
        
        \Comment $\mathsf{X}_t^{m}=\left\{x_{g,t}^{*},S_{s,t}^{*},S_{c_m,t}^{*}, P_{c_l,t}^{*}\right\}, \, \forall t\in \mathcal{T} $
        \State Calculate $\mathrm{gSCR}_{t}^m \rvert_{\mathsf{X}_t^{m}}$ and $\widehat{\mathrm{gSCR}}_{t}^m \rvert_{\mathsf{X}_t^{m}}$
        \State Update $\Delta \Omega = \bigg\{ \mathsf{X}_t^m, \mathrm{gSCR}_{t}^m \rvert_{\mathsf{X}_t^{m}} \bigg| $
        
        \hfill $\eqref{coef_ctr2}\,\mathrm{or}\,\eqref{coef_ctr3}  \mathrm{\,is\,violated} \bigg\}$
        \If{$m = 0$} 
        \State $\Omega^{m} = \bigg\{ \mathsf{X}_t^m, \mathrm{gSCR}_{t}^m \rvert_{\mathsf{X}_t^{m}} \bigg| \forall t\in\mathcal{T} \bigg\}$
        \Else
        \State $\Omega^{m} = \Omega^{m-1} \cup \Delta \Omega$
        \EndIf 
        
        \State Solve \eqref{DM3} with $\Omega^m$,
        \Comment $\mathcal{K}^{m+1}$
        \State $m = m + 1$
	\EndWhile
	\State Return $\mathcal{K}^{m}$
\end{algorithmic}
\end{algorithm}

\subsection{Power Balance and Power Flow Constraints}
\begin{subequations}
\label{PBPF}
\begin{align}
    & p_{t,n,i}^G = \sum_{\Omega_{g-i}} p_{t,n,g} + \sum_{\Omega_{w-i}} p_{t,n,w} \nonumber\\
    &  \quad\quad\quad\quad\quad + \sum_{\Omega_{m-i}} p_{t,n,m} + \sum_{\Omega_{c_m-i}} p_{t,n,c_m}, \;\;\;\;\; \forall t,n,i \label{6-}\\
    & p_{t,n,i}^D = \sum_{\Omega_{l-i}} p_{t,n,l} - \sum_{\Omega_{l-i}} p_{t,n,l}^c, \;\;\;\;\; \forall t,n,i \label{7-}\\
    & p_{t,n,i}^G - p_{t,n,i}^D = \sum_{ij\in\mathcal{R}}p_{t,n,ij}, \;\;\;\;\; \forall t,n,i\label{8-}\\
    & p_{t,n,ij} = Y_{ij}(\theta_i-\theta_j), \;\;\;\;\; \forall ij\in \mathcal{R}, t,n \label{-8}\\
    & p_{t,n,ij} \le P_{\mathrm{max},ij}, \;\;\;\;\; \forall ij\in \mathcal{R}, t,n \label{10}
\end{align}
\end{subequations}
Total active power generation $p_{t,n,i}^G$ and load $p_{t,n,i}^D$ at each time step $t$, scenario $n$, and bus $i$ are defined in \eqref{6-} and \eqref{7-} with $g/w/m/c_m/l\in \Omega_{g/w/m/c_m/l-i}$ being the set of synchronous/wind/PV/storage units/loads connected to bus $i$. Power balance at each bus is given by \eqref{8-} to \eqref{-8} where $p_{t,n,ij}$ is power flow from bus $i$ to $j$ and $ij \in \mathcal{R}$ is the set of branches; $Y$ denotes the admittance matrix of the system; $\theta_{i/j}$ is the voltage angle of bus $i/j$. Equation \eqref{10} is the line rating $P_{\mathrm{max},ij}$ constraint. \textcolor{black}{It should be noted that the DC power flow is a common simplification with acceptable performance in transmission systems for both system scheduling (unit commitment) and system planning, due to the small $R/X$ ratio. Moreover, since the focus of the proposed approach does not lie on the reactive power and system voltages, we believe this is a reasonable simplification. }

\textcolor{black}{As for the scarcity of reactive power during normal operation, one of the main reasons is that most of the current GFL converters are not required to provide reactive power and voltage support. However, the total reactive power reserves in a high IBR-penetrated system may not be reduced significantly, since the IBRs' capability of reactive power provision during normal operation is comparable to that of SGs, both of which are limited by their ratings. Enabling the IBRs to provide reactive power to help the system maintain the voltages during normal operation requires additional control/operation strategies, such as grid-forming control \cite{rosso2021grid,lasseter2019grid} and reactive power dispatch/market participation \cite{sarkar2018reactive,potter2023reactive}. As a result, in the proposed SC and BESS planning model their impact on the system reactive power and voltage during normal operation is not considered. However, if those issues also need to be considered in a specific system, linearized or conic AC power flow can be included in the proposed model \cite{taylor2012conic}.}

\subsection{Thermal Unit Constraints}
\begin{subequations}
    \begin{align}
        & \color{black} y_{t,n,g} = y_{t-1,n,g} + y_{t,n,g}^{sg} - y_{t,n,g}^{sd}, \;\;\;\;\; \forall g,t,n \label{thermal1}\\
        & \color{black} y_{t,n,g}^{sg} = y_{t,n,g}^{st}(t-T_g^{st}), \;\;\;\;\; \forall g,t,n \label{thermal2}\\
        & \color{black} y_{t,n,g}^{st} \le (1-y_{t-1,n,g}) - \sum_{j = t-T_g^{mdt}}^t y_{j,n,g}^{sd} \label{thermal3}\\
        & \color{black} y_{t,n,g}^{sd} \le y_{t-1,n,g} - \sum_{j = t-T_g^{mut}}^t y_{j,n,g}^{sg} \label{thermal4}\\
        & y_{t,n,g}  p_{\mathrm{min},g} \le p_{t,n,g} \le y_{t,n,g} p_{\mathrm{max},g}, \;\;\;\;\; \forall g,t,n \label{2-} \\
        & -R_d \le  p_{t,n,g} - p_{t-1,n,g} \le R_u , \;\;\;\;\; \forall g,t \label{ramp}
    \end{align}
\end{subequations}
\textcolor{black}{The operation of thermal units are modeled in \eqref{thermal1}-\eqref{thermal4}. Constraint \eqref{thermal1} defines the commitment state of each generator as ‘online' if the unit was already generating in the previous period unless it has been shut down in the current period, or has started generating in the current period. Constraint \eqref{thermal2} sets the state of a unit as ‘starts generating’ in the current time period if the unit was started up $T^{st}_g$ periods earlier. A unit can only be started up if it has been offline for at least $T^{mdt}_g $ periods and was offline in the previous period, as defined by \eqref{thermal3}. Finally, \eqref{thermal4} enforces that a generator can only be shut down if it was online in the previous period, as well as having been online for at least $T^{mut}_g$ periods.} Active power generation of thermal units, $p_{t,n,g}$ is bounded by their minimum and maximum limits ($p_{\mathrm{max},g}$ and $p_{\mathrm{min},g}$) as in \eqref{2-}. The ramp constraint of the thermal units is considered in \eqref{ramp} with $R_d$ and $R_u$ being the ramp down and up limits.

\subsection{Constraints of battery storage system} \label{sec:4.5}
\begin{subequations}
\label{eq:storage}
\begin{align}
    & p_{t,s,c_m} = p_{t,s,c_m}^{\mathrm{dch}} - p_{t,s,c_m}^{\mathrm{ch}}, \;\;\;\;\; \forall\, t,s,c_m 
    \label{p_b}\\
    & \color{black} 0 \le p_{t,s,c_m}^{\mathrm{dch}} \le \bar P_{c_m}^{\mathrm{dch}} v_{c_m}, \;\;\;\;\; \forall\, t,s,c_m \label{dis1} \\
    & \color{black} 0 \le p_{t,s,c_m}^{\mathrm{ch}} \le - \bar P_{c_m}^{\mathrm{ch}} (1-v_{c_m}), \;\;\;\;\; \forall\, t,s,c_m \label{dis2}, \\
    &\mathrm{SoC}_{t,s,c_m}S_{c_m} = \mathrm{SoC}_{t-1,s,c_m}S_{c_m} - \frac{1}{\eta_{c_m}} p_{t,s,c_m}^{\mathrm{dch}}\Delta t \nonumber \\
    \label{soc_cal}
    & + {\eta_{c_m}} p_{t,s,c_m}^{\mathrm{ch}}\Delta t, \;\;\;\;\; \forall\, t,s,c_m \\
    \label{soc_lim}
    & \mathrm{SoC}_{\mathrm{min}} \le \mathrm{SoC}_{t,s,c_m} \le \mathrm{SoC}_{\mathrm{max}}, \;\;\;\;\; \forall\, t,s,c_m\\
    \label{SoC_T}
    & \mathrm{SoC}_{0,s,c_m} = \mathrm{SoC}_{T,s,c_m}, \;\;\;\;\; \forall\, s,c_m.
\end{align}
\end{subequations}
The power injection from BESS to the \textcolor{black}{grid} ($p_{t,s,c_m}$) is confined in \eqref{p_b}-\eqref{dis2} by the upper bound of the charging ($\bar P_{c_m}^{\mathrm{ch}}$) and discharging ($\bar P_{c_m}^{\mathrm{dch}}$) rate. \textcolor{black}{The binary variable $v_{c_m}$ ensures the BESS does not charge and discharge at the same time.} The battery state of charge ($\mathrm{SoC}_{t,s,c_m}$) is quantified by \eqref{soc_cal} with the charging/discharging efficiency $\eta_{c_m}$. \eqref{soc_lim} imposes the upper ($\mathrm{SoC}_{\mathrm{max}}$) and lower ($\mathrm{SoC}_{\mathrm{min}}$) limits on the SoC of the storage devices. The SoC at the end of the considered time horizon ($t=T$) is set to be a pre-specified value equal to its initial value as in \eqref{SoC_T}. \textcolor{black}{In summary, the system planning model contains the objective function \eqref{eq:SUC} and the constraints \eqref{SCC_linear}, \eqref{ESCR_linear}, \eqref{PBPF}-\eqref{eq:storage}.}

\textcolor{black}{Notably, it is not realistic to guarantee system stability during the system planning stage. The proposed model and the incorporation of the developed constraints are to ensure there are enough resources in the future system to maintain system stability. During day-ahead operation planning with more accurate forecasting, the stability constraints can also be included in the system scheduling model such as unit commitment and economic dispatch problems, to ensure system stability. Finally, in near real-time, the stability assessment based on dynamic simulation with more detailed models can always be carried out as the last resort to guarantee system stability.}

\begin{figure}[!t]
    \centering
	\scalebox{1}{\includegraphics[trim=0 0 0 0,clip]{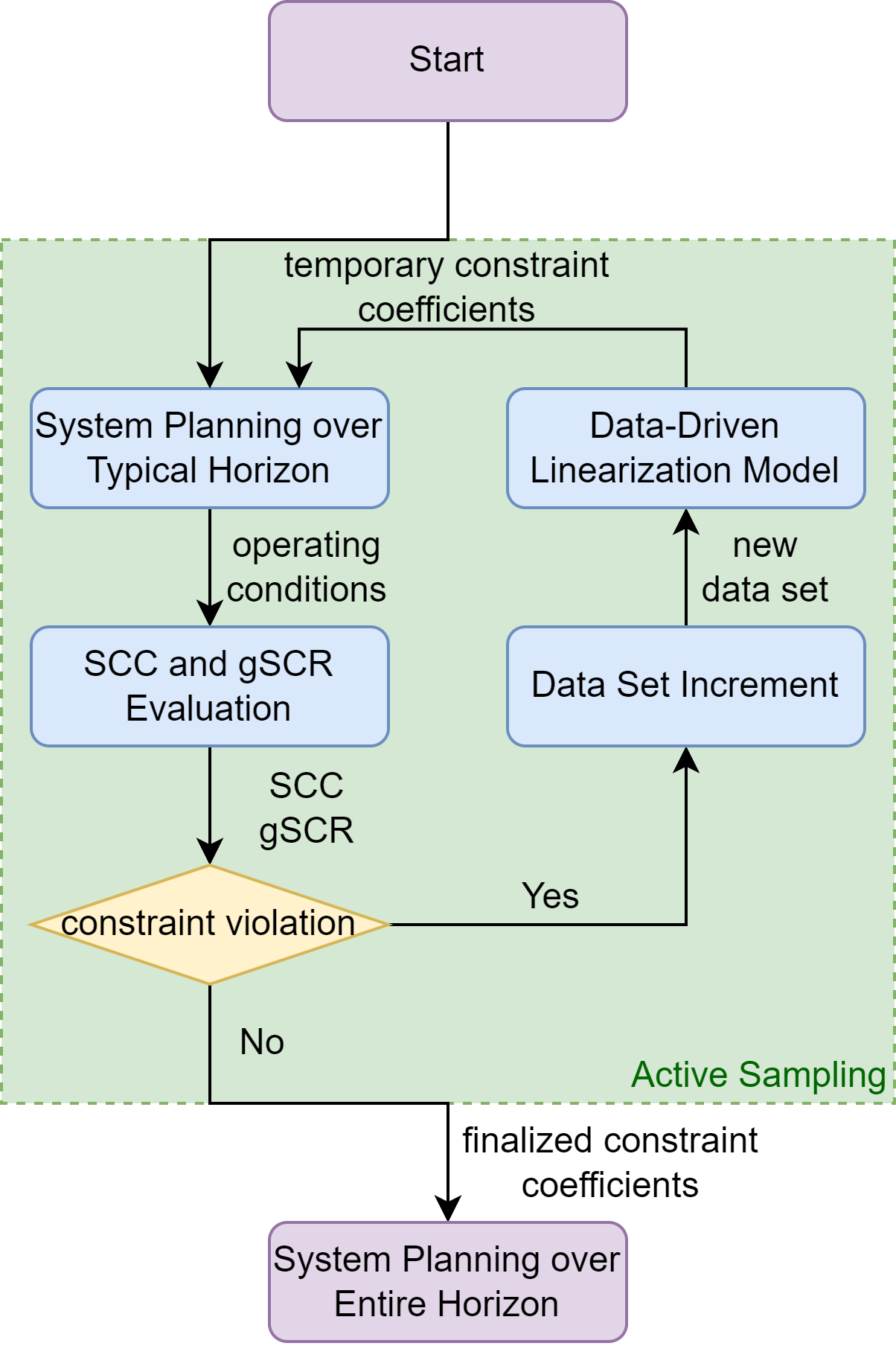}}
    \caption{\label{fig:diagram}Overall framework of the proposed method.}
\end{figure}

\textcolor{black}{The overall framework of the proposed model is depicted in Fig.~\ref{fig:diagram}. The SCC and gSCR constraint coefficients for the system planning model are determined utilizing the data-driven linearization approach \eqref{DM3} and \eqref{Omega} based on a representative data set. This data set is generated within the proposed active sampling method where the data set in each iteration is increased by combining the previous data set and the new samplings whose SCC and gSCR values do not meet the requirements according to the current SCC and gSCR constraint coefficients. The performance of the active sampling approach is demonstrated in Section~\ref{sec:5.1.2}.}

\section{Case Studies} \label{sec:5}
\begin{figure}[!t]
    \centering
    \vspace{-0.4cm}
	\scalebox{0.48}{\includegraphics[trim=0 0 0 0,clip]{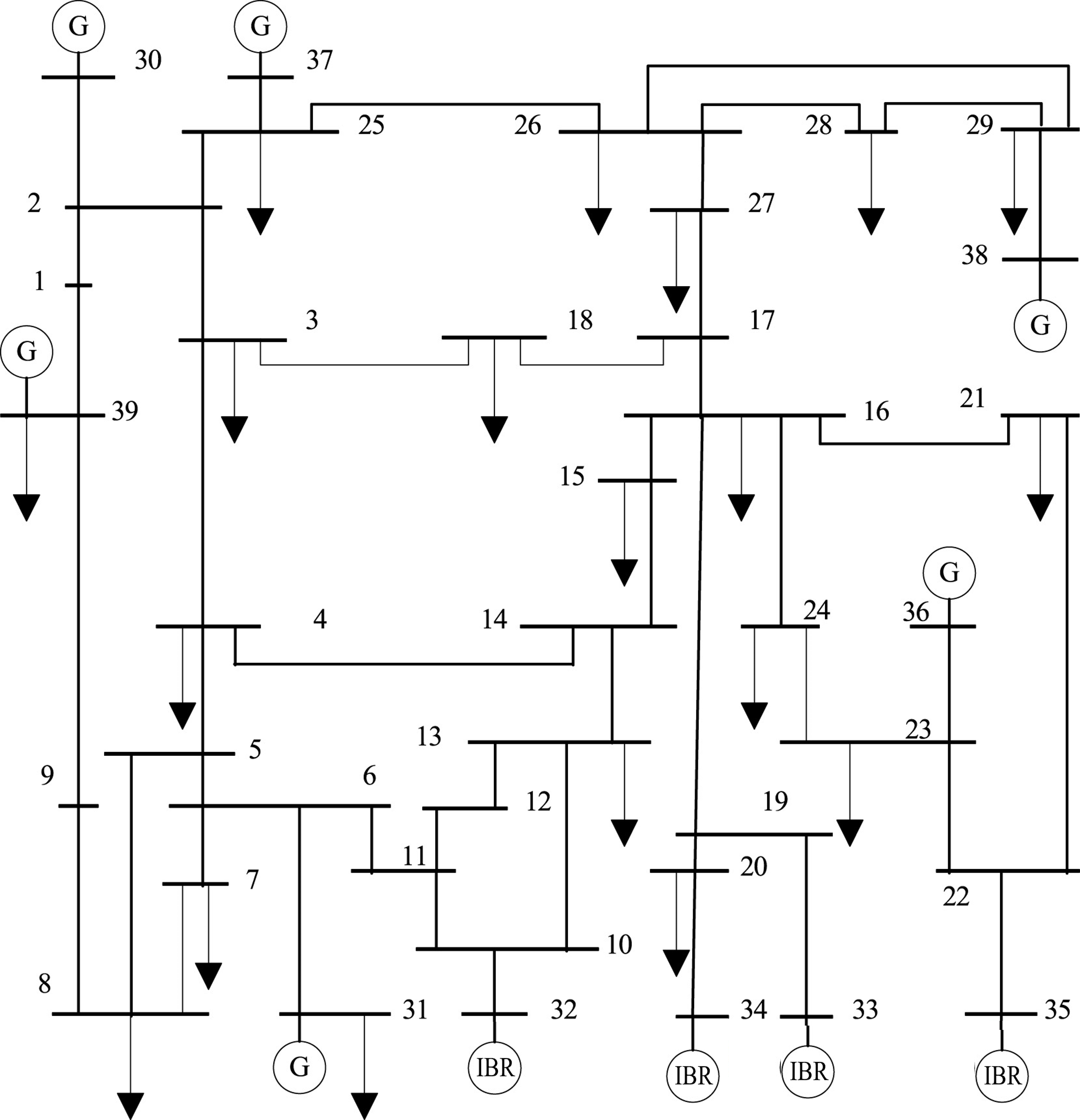}}
    \caption{\label{fig:39-bus}Modified IEEE-39 bus system.}
    \vspace{-0.4cm}
\end{figure} 
The IEEE-39 bus system shown in Fig.~\ref{fig:39-bus} is utilized as the test system for constraints validation and applications. The models and parameters related to the test system are introduced in this section. To increase the renewable penetration, IBRs are added at Bus 32, 33, 34, and 35.

The parameters of transmission lines and loads are available in \cite{39_bus}. The load and renewable generation profile in \cite{6026941,9968474} is adapted for the simulation during the considered time horizon. The characteristics of thermal generators are given in Table~\ref{tab:SG_para} while considering the data in \cite{9403907,9968474}, with the location of the three types being Bus $\{30,37\}$, $\{31,36,38\}$ and $\{39\}$ respectively. Other system parameters are set as follows: load demand $P^D\in [5.16, 6.24]\,\mathrm{GW}$, base power $S_B = 100\mathrm{MVA}$, \textcolor{black}{load shedding cost $c^{VOLL} = 30\,\mathrm{k\pounds/MWh}$}. \textcolor{black}{Note that the VOLL is in general difficult to estimate in practice and may vary depending on many factors. However, since we only focus on enhancing the system stability through additional assets, i.e., SCs and BESSs, and assume that there will be enough generation to meet the demand in the system. Therefore, the VOLL has little impact on the proposed model.} The annualized investment costs are $1.84\,\mathrm{k\pounds/MVA\cdot yr}$ for SCs \cite{mahmud2022improvement} and  BESS for $19.88\,\mathrm{k\pounds/MW\cdot yr}$ \cite{augustine2021storage}. \textcolor{black}{Although the proposed MILP-based optimization problem is nonconvex by definition, due to the existence of integer variables, it can still be efficiently solved by applying the branch-and-bound algorithm with the optimality being ensured with an MIP gap of 0.5\% \cite{MIP}.}


\begin{table}[!t]
\renewcommand{\arraystretch}{1.2}
\caption{{Parameters of Thermal Units}}
\label{tab:SG_para}
\noindent
\centering
    \begin{minipage}{\linewidth} 
    \renewcommand\footnoterule{\vspace*{-5pt}} 
    \begin{center}
        \begin{tabular}{ c || c | c | c }
            \toprule
            Type & Type I & Type II & Type III \\ 
            \cline{1-4}
            No-load Cost [k\pounds/h]& 4.5 & 3 & 0\\
            \cline{1-4} 
            Marginal Cost [\pounds/MWh]& 47 & 200 & 10\\
            \cline{1-4} 
            Startup Cost [k\pounds]& 10 & 0 & N/A \\
            \cline{1-4} 
            Startup Time [h]& 4 & 0 & N/A\\
            \cline{1-4}
            Min Up Time [h] & 4 & 0 & N/A\\
            \cline{1-4} 
            Min Down Time [h] & 1 & 0 & N/A \\
            \cline{1-4}
            Inertia Constant [s]& 6 & 6 & 6 \\
           \bottomrule
        \end{tabular}
    \end{center}
    \end{minipage}
\end{table} 

\subsection{Model validations}

\subsubsection{SCC convergence}

\begin{figure}[!b]
    \centering
    \vspace{-0.4cm}
	\scalebox{1.2}{\includegraphics[trim=0 0 0 0,clip]{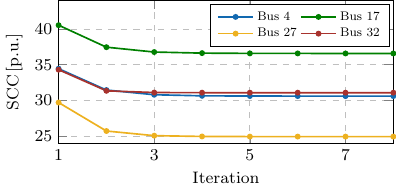}}
    \caption{\label{fig:SCC_Convergence} Convergence of SCC calculation.}
\end{figure}

To demonstrate the convergence of the SCC calculation method proposed in Algorithm~\ref{alg:SCC}, the SCC values in each iteration are reported in Fig.~\ref{fig:SCC_Convergence}. Note that only the SCCs at four buses selected arbitrarily are shown in the figure for clarity and those at other buses present a similar trend. It can be observed that the initial SCCs at different buses have the highest value. This is because the terminal voltages of the IBRs are initialized by $\Delta V^{(0)}_{\Phi(c)} = -1$ in Algorithm~\ref{alg:SCC}, thus leading to the highest current injection from IBR according to the droop control. A fast convergence of the SCC can also be observed within a few iterations, demonstrating the effectiveness of the proposed algorithm. The decreasing trend is due to the declined voltage deviation in each iteration, i.e., $\left|\Delta V^{(k)}_{\Phi(c)}\right| \ge \left|\Delta V^{(k+1)}_{\Phi(c)}\right|$.
\subsubsection{Effectiveness of active sampling} \label{sec:5.1.2}

\begin{figure}[!t]
    \centering
	\scalebox{1.16}{\includegraphics[trim=0 0 0 0,clip]{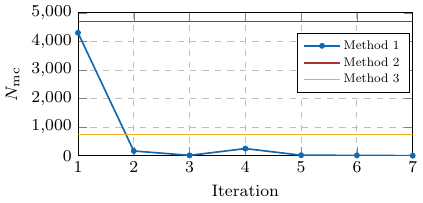}}
    \caption{\label{fig:Sampling} Convergence of active sampling.}
\end{figure}
The performance of constraint linearization with the active sampling algorithm as discussed in Section~\ref{sec:4.2.3} and \ref{sec:4.2.4} is assessed here. The results are shown in Fig.~\ref{fig:Sampling} with $N_\mathrm{mc}$ being the number of misclassifications during one-year operation (8760 hours). Note that since in the proposed linearization, most of the linearized values are only required to be at the same sides as the true data samples, as defined in \eqref{coef_ctr2} and \eqref{coef_ctr3}, the absolute errors are of no concern and not considered here.

For comparison, three different methods are considered, where Method 1 is the proposed linearization \eqref{DM3}, with the active sampling; Method 2 is the Least Squares Regression (LSR) with random sampling; Method 3 is the proposed linearization \eqref{DM3}, with random sampling. Methods 2 and 3 require no iteration, hence being flat curves. It can be observed from Fig.~\ref{fig:Sampling} that Method 2 presents the highest number of misclassifications (around half of the hours) since the LSR penalized the under- and over-estimation equally. This is improved with the proposed boundary-aware linearization approach \eqref{DM3}, as indicated by the yellow line. However, the actual operating conditions are not considered in the sampling process, resulting in a significant number of misclassifications, which endangers system operation. As for Method 1, the number of misclassifications decreases fast within a few iterations, demonstrating the effectiveness of the proposed linearization and active sampling approach.

\subsection{Benefit of BESS and SC coordinated planning}
\textcolor{black}{Since most of the research in the area of BESS placement and sizing does not consider the stability constraints, e.g.,  \cite{bahramirad2012reliability, chen2012sizing}, we define it as the Base Case together with the following cases to demonstrate the benefit of the proposed model.}

\begin{itemize}
    \item Base Case: Stability constraints not considered during planning, but maintained at the operation stage.
    \item Case I: Coordinated planning of BESS and SC. 
    \item Case II: Optimal planning of BESS only. 
    \item Case III: Optimal planning of SC only.  
\end{itemize}
\begin{table}[!b]
\renewcommand{\arraystretch}{1.2}
\caption{{Investment and operational cost}}
\label{tab:Inv_Cost}
\noindent
\centering
    \begin{minipage}{\linewidth} 
    \renewcommand\footnoterule{\vspace*{-5pt}} 
    \begin{center}
        \begin{tabular}{ c || c | c | c | c | c | c}
            \toprule
            & \multicolumn{2}{c|}{BESS Investment} & \multicolumn{2}{c|}{SC Investment} & Ope. & Total \\
            \cline{2-5}
        Case & Cap.      & Cost    & Cap.      & Cost    & Cost & Cost \\
 &      $[\mathrm{GW}]$        &   $[\mathrm{m\pounds/y}]$      &     $[\mathrm{GW}]$  & $[\mathrm{m\pounds/y}]$ &  $[\mathrm{k\pounds/h}]$ &  $[\mathrm{m\pounds/y}]$\\
\cline{1-7}
Base   &       $2.93$       &    $58.19$     &          $0$     &  $0$  &  $181.91$  &  $1651.72$\\
\cline{1-7}
I   &       $2.63$       &    $52.23$     &          $0.38$     &  $0.70$  &  $165.63$ &  $1503.85$ \\
\cline{1-7}
II  &       $3.96$       &    $78.68$     &        $0$       & $0$ & $174.94$  &  $1611.15$  \\
\cline{1-7}
III &      $0$       &    $0$     &          $1.06$     &  $1.96$  &  $440.18$  & $3857.94$  \\
           \bottomrule
        \end{tabular}
    \end{center}
    \end{minipage}
\end{table} 
The system planning model of one year is solved for each case defined above. The results are listed in Table.~\ref{tab:Inv_Cost} with the investment decisions and costs of both BESSs and SCs as well as the system operation cost considered. Note that the investment cost is the annualized prices whereas the system operation (Ope.) cost is the averaged hourly operational cost. \textcolor{black}{In Base Case, the system strength and SCC constraints are not considered during the system planning stage and a second operation model is run with the investment decisions being fixed according to the solution of the planning model. As a result, the optimal investment decision only contains BESS. The stability constraints are further maintained at the operation stage by committing more SGs, thus leading to more operational and total costs compared with Case I and II.} Our proposed model (Case I) provides the lowest operation cost due to the coordination of BESSs and SCs. A total investment of $2.63\,\mathrm{GW}$ BESSs with the cost of $52.23\,\mathrm{M\pounds/yr}$ and $0.38\,\mathrm{GW}$ SCs with the cost of $0.70\,\mathrm{M\pounds/yr}$ are needed. The BESSs are planned to provide the energy and power shifting capability to the system such that more renewable resources can be utilized. These BESSs at the same time provide a certain amount of system strength and SCC, which is however not sufficient to maintain the system requirements. Additional system strength and SCC are provided by the SCs, due to their much lower investment cost and higher SCC capacity compared with BESS. 

In Case II, since only BESSs are available during the planning stage, to provide the same amount of system strength and SCC that are supplied by $0.38\,\mathrm{GW}$ SCs in Case I, much more BESSs ($1.33\,\mathrm{GW}$) at planning stage and additional SGs at operation stage are required due to the lower SCC capacity of BESSs, thus inducing more investment cost and operational cost respectively compared with Case I. In Case III where only the SCs are available, a lower total investment capacity of $1.06\,\mathrm{GW}$ is planned compared with Case I since the SCs are only needed to enhance the system strength and provide SCC. However, without any storage in the system, the operational cost is much higher than the previous two cases, as a significant amount of renewable energy is curtailed. Therefore, it is clear that the proposed approach where the BESSs and SCs can be coordinated achieves the lowest investment and operation cost.

\subsection{\textcolor{black}{Importance of Considering Both System Strength and SCC Constraints}}
The necessity of including both system strength and SCC constraints in the system planning model is demonstrated here. The results are shown in Table~\ref{tab:two_contrs} where three combinations of the constraints are considered with $\mathrm{SS}=0/1$ and $\mathrm{SCC}=0/1$ being to solve the planning model without/with the system strength and SCC constraints respectively. In the case where only the SCC constraint is considered, the investment decisions are very similar to the case where both constraints are included. Specifically, the decision for the SC is almost the same for the purpose of SCC provision. Slightly more BESSs are invested to decrease the operational cost. However, the system strength constraints are violated for $0.84\%$ of the time within one-year operation. Although this number is small due to the large amount of BESS operated in GFM, it still endangers system operation. As for the case where only the system strength constraint is considered, the investment in SC is significantly reduced since the SCC is no longer enforced by the constraints. As a result, the SCC constraints are violated for more than half of the time in spite of the decreased operational cost, which demonstrates the importance of including both system strength and SCC constraints in the system planning model. 

\begin{table}[!b]
\renewcommand{\arraystretch}{1.2}
\caption{{Impact of system strength and SCC constraints}}
\label{tab:two_contrs}
\noindent
\centering
    \begin{minipage}{\linewidth} 
    \renewcommand\footnoterule{\vspace*{-5pt}} 
    \begin{center}
        \begin{tabular}{ c || c | c | c | c | c }
            \toprule
           Constraint & \multicolumn{2}{c|}{Investment} & \multicolumn{2}{c|}{Violation Rate} &{Ope.} \\
            \cline{2-5}
       [SS,  & BESS      & SC    & SS      & SCC    & Cost \\
SCC] &      $[\mathrm{GW}]$        &   $[\mathrm{GW}]$      &     $[\mathrm{\%}]$  & $[\%]$ &  $[\mathrm{k\pounds/h}]$\\
\cline{1-6}
[1,1]  &       $2.63$       &    $0.38$     &          $0$     &  $0$  &  $165.63$ \\
\cline{1-6}
[0,1]  &       $2.72$       &    $0.36$     &        $0.84$       & $0$ & $163.51$    \\
\cline{1-6}
[1,0] &      $2.64$       &    $0.02$     &          $0$     &  $54.58$  &  $135.30$    \\
           \bottomrule
        \end{tabular}
    \end{center}
    \end{minipage}
\end{table}

\subsection{Impact of wind penetration}
The investment situations with various installed wind capacities in different cases are investigated in this section to demonstrate the importance of the proposed coordinated planning at different IBR penetrations. However, as the BESS is not available in Case III, the investment cost is much lower whereas the operation cost is much higher than the other cases as explained in the previous section, hence not being considered in this section. Instead, the following case is defined where the BESSs and SCs planning are decoupled, representing an optimal design of the system where the BESSs are optimally placed for power balance only and the SCs are further placed to maintain the system strength and SCC constraints given the already-determined BESSs. 
\begin{itemize}
    \item Case IV: The optimal investment of BESSs is first determined by running the planning model without the system strength and SCC requirements, which are further maintained with SCs by a second run. 
\end{itemize} 

\begin{figure}[!t]
    \centering
    \vspace{-0.4cm}
	\scalebox{1.2}{\includegraphics[trim=0 0 0 0,clip]{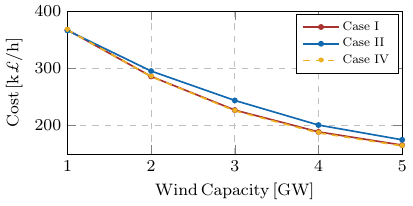}}
    \caption{\label{fig:Cost1} System operation cost with different investment strategies.}
    \vspace{-0.4cm}
\end{figure}

\begin{figure}[!b]
    \centering
    \vspace{-0.4cm}
	\scalebox{1.2}{\includegraphics[trim=0 0 0 0,clip]{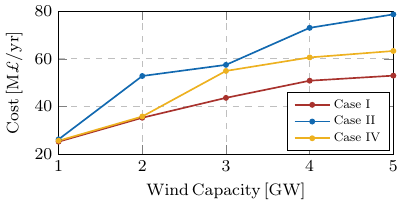}}
    \caption{\label{fig:Cost2} Investment cost with different investment strategies.}
\end{figure}

The results are depicted in Fig.~\ref{fig:Cost1} (system operation cost) and Fig.~\ref{fig:Cost2} (investment cost). It is clear from Fig.~\ref{fig:Cost1} that as the wind capacity increases, the operation cost in all the cases decreases, since with proper amounts of BESSs in the system, the increased wind power can be effectively utilized. The operational cost in Case II is slightly higher than the other cases since more SGs are dispatched online to provide system strength and SCC due to the high investment cost of BESSs. The operation costs in Case I and IV are almost identical to each other as they both have sufficient amounts of BESSs and SCs for power balance and system strength, SCC provision.

As for the investment cost (Fig.~\ref{fig:Cost2}), an increasing trend can be observed along the growth of the wind capacity and Case II always presents the highest cost due to the BESS-only investment strategy. Moreover, the investment cost of Case IV becomes higher than that of the proposed method (Case I) at higher wind penetration, this is because with the decoupled planning, the BESS placement in Case IV does not consider the system strength and SCC requirement, and some redundancy in the BESS and SC capacity is resulted in, thus leading to more investment cost compared with the proposed method. 

\subsection{Sensitivity Analysis of IBR Overloading Capability and SCC Requirement}
\begin{figure}[!t]
    \centering
    \vspace{-0.4cm}
	\scalebox{1.16}{\includegraphics[trim=0 0 0 0,clip]{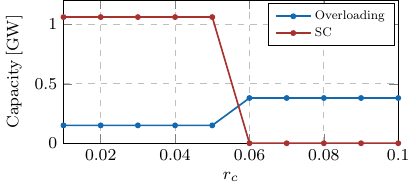}}
    \caption{\label{fig:Overloading} Investment capacity of SC and IBR overloading.}
\end{figure}

In previous sections, the temporary overloading capability of IBR is set as $1.2\,\mathrm{p.u.}$, which is much less than that of an SG/SC. As mentioned in the introduction, different approaches that can achieve a higher overloading capability of IBRs have been proposed. Therefore, the optimal planning decision may be a combination of BESS, SC, and the additional temporary overloading capability of IBRs, depending on their investment cost. However, different from the costs of BESS and SC, which are available in the literature and technical reports, the development of IBR temporary overloading capability is still in the research stage with the investment costs being unclear and varying significantly depending on the deployed technologies.

\begin{table}[!b]
\renewcommand{\arraystretch}{1.2}
\caption{Impact of SCC requirement levels}
\label{tab:SCC_lim}
\noindent
\centering
    \begin{minipage}{\linewidth} 
    \renewcommand\footnoterule{\vspace*{-5pt}} 
    \begin{center}
        \begin{tabular}{ c || c | c | c | c | c }
            \toprule
           SCC & \multicolumn{2}{c|}{BESS Investment} & \multicolumn{2}{c|}{SC Investment} & Ope. \\
            \cline{2-5}
        limit & Cap.      & Cost    & Cap.      & Cost    & Cost \\
factor &      $[\mathrm{GW}]$        &   $[\mathrm{M\pounds/yr}]$      &     $[\mathrm{GW}]$  & $[\mathrm{M\pounds/yr}]$ &  $[\mathrm{k\pounds/h}]$\\
\cline{1-6}
1.0 &       $2.63$       &    $52.23$     &          $0.38$     &  $0.70$  &  $165.63$ \\
\cline{1-6}
1.3  &       $2.62$       &    $52.08$     &        $1.13$       & $2.09$ & $165.79$    \\
\cline{1-6}
1.6 &      $2.62$       &    $52.08$     &          $2.03$     &  $3.75$  &  $165.70$    \\
           \bottomrule
        \end{tabular}
    \end{center}
    \end{minipage}
\end{table} 

Therefore, the impact of the IBR temporary overloading cost on the optimal investment decisions is investigated with the results plotted in Fig.~\ref{fig:Overloading} where $r_c$ is defined as the ratio of IBR temporary overloading cost to the IBR permanent overloading cost (i.e., increasing the BESSs' converter capacity). It can be observed in the figure that when $r_c$ is greater than $0.06$, no IBR temporary overloading capacity is planned. Instead, SCs are utilized to supply SCC in the system since SCs are cheaper for the same amount of SCC provision. However, if $r_c$ becomes smaller than $0.06$, a certain amount of IBR temporary overloading capacity ($1.06\,\mathrm{GW}$) would be more beneficial to the system, which reduces the SC capacity from $0.38\,\mathrm{GW}$ to $0.15\,\mathrm{GW}$. Note that the SC capacity is not reduced to zero since it also provides system strength which cannot be achieved by IBR temporary overloading. Moreover, the capacity of IBR temporary overloading required for SCC provision is much higher than the SC capacity being replaced $0.23\,\mathrm{GW}$, indicating SCs provide more SCC than the IBR given the same capacity.

The impact of different levels of SCC requirements is also assessed here with the results listed in Table~\ref{tab:SCC_lim}, where the SCC limit factor is the ratio of the SCC limit to that in Case I. It can be observed that compared with Case I, as the SCC limit increases, the BESS investment decisions are barely influenced, since their main roles in the system are power/energy balancing and system strength improvement. On the contrary, the invested SC capacities increase in an approximately linear fashion to maintain the increased SCC requirement, due to their superior SCC provision capability.


\subsection{\textcolor{black}{Scalability of Proposed Model}}
The computational performance and the stability of the proposed model are demonstrated through IEEE 118-bus system with the demand $[2000,5600]\,\mathrm{MW}$ and $S_B = 1000\,\mathrm{MVA}$. Wind generation is added at Bus 10, 26, 49, 61, 65 and 80 with a total capacity of $6\,\mathrm{GW}$. The generator parameters remain unchanged. The network data of the systems can be obtained in \cite{Data_system}. The results are shown in the table below. The ``D'' and ``S'' represent deterministic and stochastic defined by a scenario tree with one and nine branches respectively, as indicated in the response to Comment 5. The computational time in the case, with SC is increased due to the inclusion of the Stability Constraints (SCtr). However, this increment is acceptable, especially in the stochastic case where the computational burden mainly comes from the nature of stochastic optimization and further incorporation of the SC does not increase the computational time significantly, thus highlighting the efficiency of the proposed method. 
    
\textcolor{black}{It can also be observed that although the computational time compared with the IEEE 39-bus system is increased, it is still within the acceptable range and the main computational burden comes from the stochastic formulation itself, which is out of the scope of the proposed model. The additional time increased due to the stability constraints is insignificant.  As for the other results, due to their similarity to those in IEEE 39-bus system, they are not presented here. However, planning a system with thousands of nodes while considering the operational details can be challenging with or without the proposed stability constraints. In this case, some simplification methods may be necessary such as zonal planning, which simplifies the system into different zones while keeping its main features and investment drivers, and the simplified system is then expanded optimally for the full scope \cite{lumbreras2017large} or scenario reduction \cite{heitsch2003scenario}.}

\begin{table}[!b]
\renewcommand{\arraystretch}{1.2}
\caption{\textcolor{black}{Computational time of different cases and systems}}
\label{tab:time}
\noindent
\centering
    \begin{minipage}{\linewidth} 
        \renewcommand\footnoterule{\vspace*{-5pt}} 
        \begin{center}
            \begin{tabular}{ c || c | c | c | c }
                \toprule
                  & \multicolumn{2}{c|}{$\mathbf{Step\,Time\,[s/step]}$} & \multicolumn{2}{c}{$\mathbf{Total\,Time\,[h]}$} \\
                 \cline{2-5}
                 & without SCtr & with SCtr & without SCtr & with SCtr  \\ 
                \cline{1-5}
                \textbf{39-bus D} & 0.155 & 0.212 & 0.377 & 0.516  \\
                \cline{1-5}
                \textbf{39-bus S} & 10.486 & 11.922 & 25.516 & 29.181  \\
                \cline{1-5}
                \textbf{118-bus D} & 0.299 & 0.453 & 0.728 & 1.102  \\
                \cline{1-5}
                \textbf{118-bus S} & 24.148 & 26.035 & 58.760 & 63.352  \\
               \bottomrule
            \end{tabular}
        \end{center}
        \end{minipage}
\end{table}

\section{Conclusion} \label{sec:6}
This paper proposes a coordinate synchronous condenser and BESS planning model for small-signal and transient stability improvement in weak grids. Sufficient system strength for small-signal stability and SCC during transient processes are ensured with minimum investment and operational costs, by optimally placing the SCs and BESSs in the system. The system strength and SCC constraints are developed considering the different characteristics of the SC and BESS. An iterative SCC calculation algorithm is proposed to account for the dependence between the IBR terminal voltage and current injection. The highly nonlinear constraints are linearized through a data-driven method where an active sampling approach is proposed to generate a representative data set. 

The effectiveness and scalability of the proposed coordinated planning model are demonstrated through case studies based on IEEE 39- and 118-bus systems. Less investment and operational costs are needed with the proposed method compared to the cases where only one resource is available and the case where the SC and BESS are planned separately. The value of the IBR temporary overloading capability is also investigated with the impact of the overloading cost on the optimal planning decisions being revealed. 

\renewcommand{\theequation}{A.\arabic{equation}}

\setcounter{equation}{0}

\section*{\textcolor{black}{Appendix A. Derivation of SCC}} \label{sec:app_SCC}

\begin{figure}[!t]
  \centering
    \begin{minipage}{0.5\textwidth}
        \centering
        \scalebox{0.68}{\includegraphics[trim={0.2cm 0cm 1.0cm .0cm}]{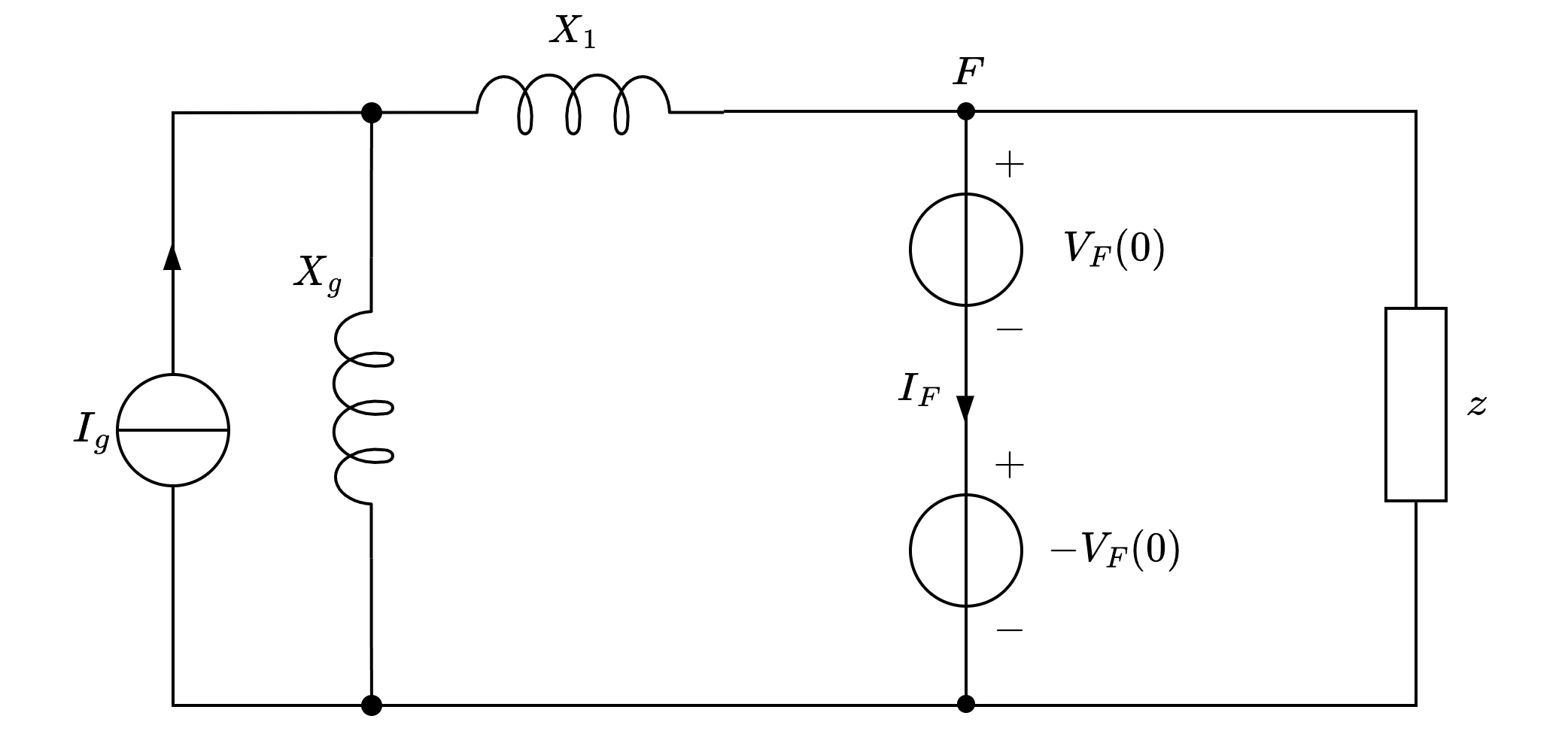}} 
        \vspace{0.15cm}
    \end{minipage} 
    \begin{minipage}{0.5\textwidth}
        \centering
        \scalebox{0.68}{\includegraphics[trim={0.2cm 0cm 0.9cm .0cm}]{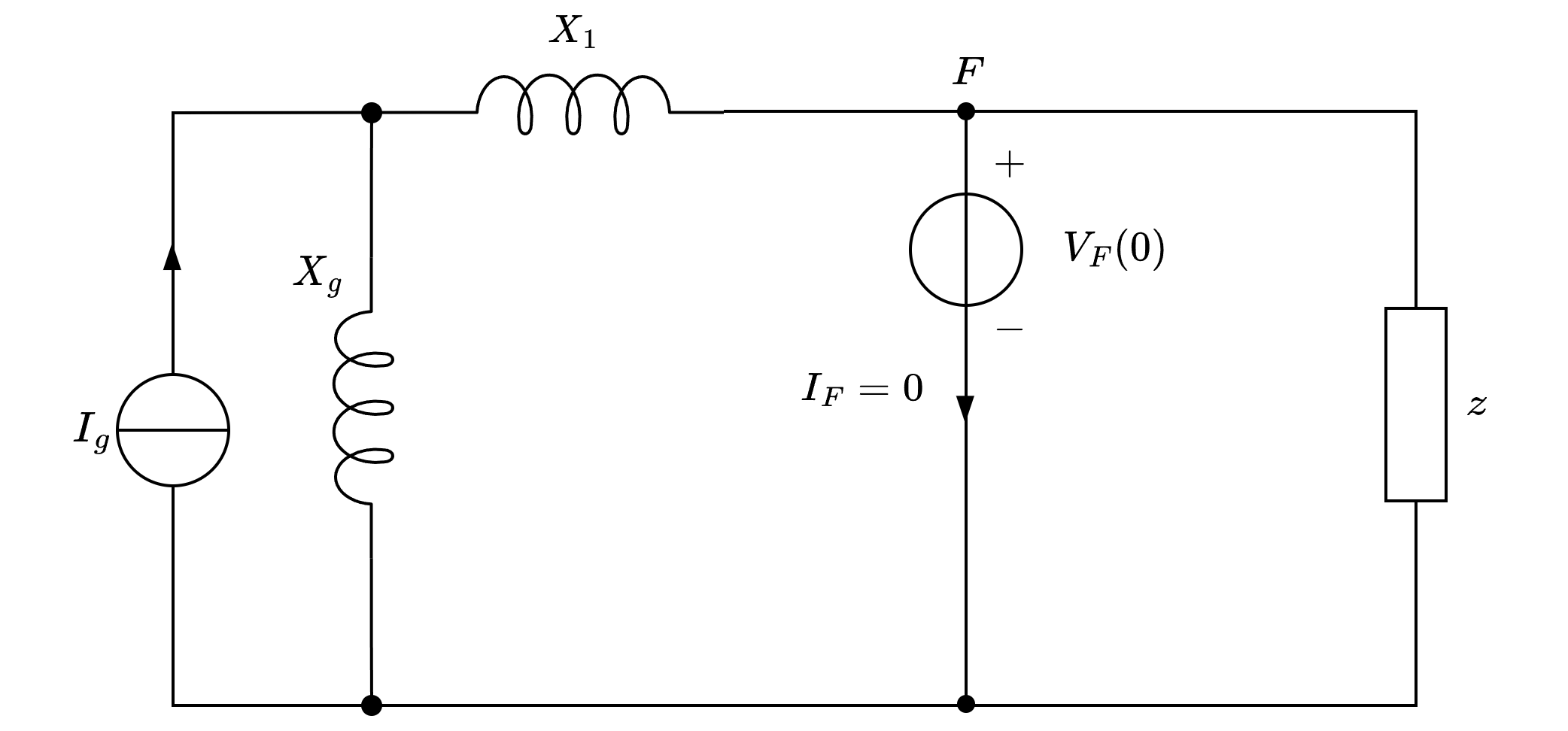}}
    \end{minipage} 
    \begin{minipage}{0.5\textwidth}
        \centering
        \scalebox{0.68}{\includegraphics[trim={0.1cm 0cm 0cm .0cm}]{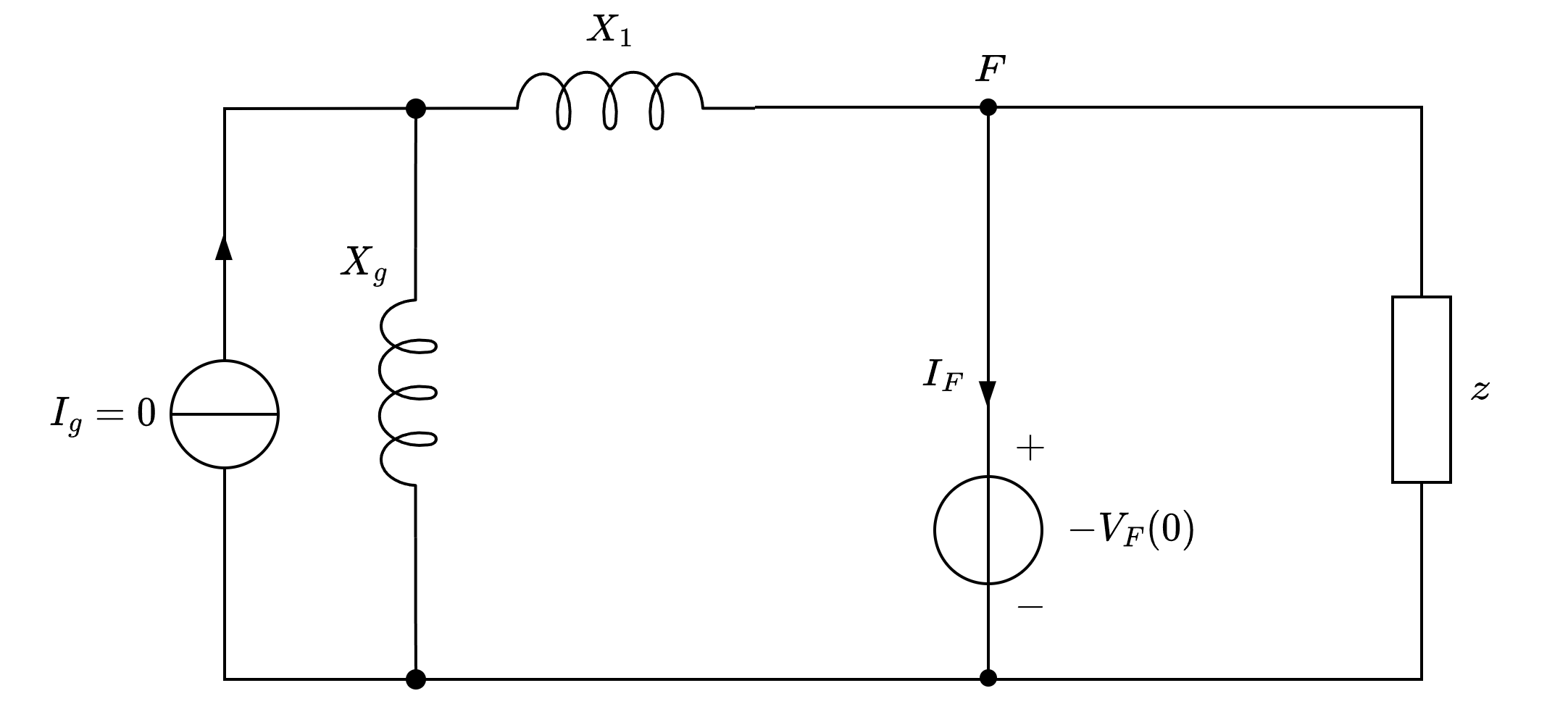}}
    \end{minipage} 
  \caption{\label{fig:SG_SC}\textcolor{black}{Equivalent circuit to calculate the SCC from SGs: (i) post-fault condition; (ii) pre-fault condition; (iii) `pure'-fault condition.}}
\end{figure}

We first briefly revisit the conventional superposition principle for the three-phase short circuit fault current calculation in an SG-dominated system. The equivalent circuit of a loaded SG during a short circuit fault at point F is shown in Fig.~\ref{fig:SG_SC}, where all the symbols have their usual meaning. According to the superposition principle, the SCC at $F$, $I_F$ can be evaluated by analyzing the `pure'-fault condition, where the only source in the system is $-V_F(0)$:
    \begin{align}    
    \label{IF_SG}
        I_F = \frac{-V_F(0)}{Z_{FF}}        
    \end{align}
where $Z_{FF}$ is the equivalent impedance seen at node $F$, i.e., $(X_{g}+X_1)//Z$ in Fig.~\ref{fig:SG_SC} or the diagonal element at node $F$ of the system impedance matrix in a general multi-machine system. Similarly, the equivalent circuit for SCC calculation with IBRs in the system is given below.

\begin{figure}[!t]
  \centering
    \begin{minipage}{0.5\textwidth}
        \centering
        \scalebox{0.68}{\includegraphics[trim={0.3cm 0cm 1.0cm .0cm}]{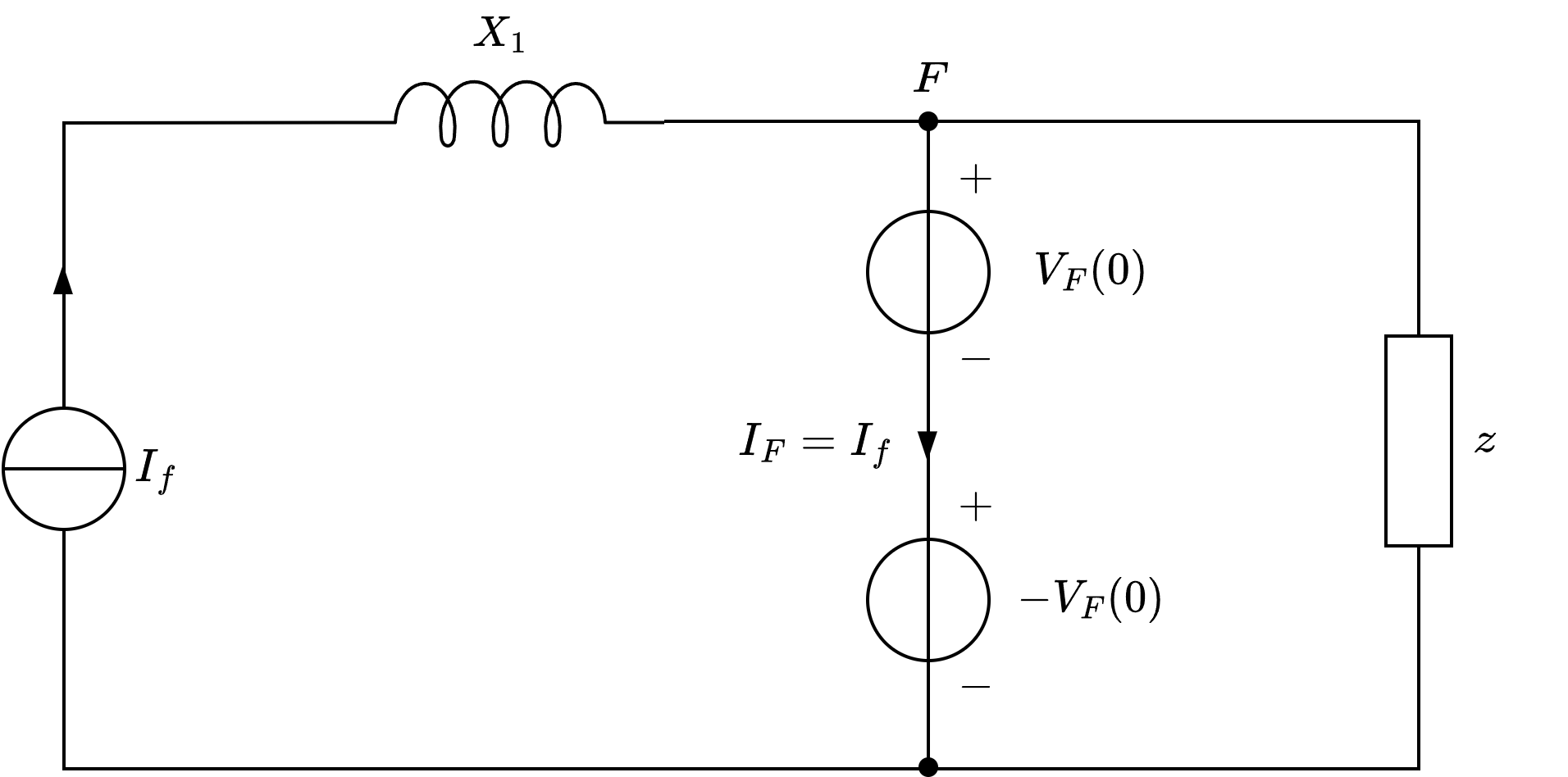}} 
        \vspace{0.15cm}
    \end{minipage} 
    \begin{minipage}{0.5\textwidth}
        \centering
        \scalebox{0.68}{\includegraphics[trim={0.3cm 0cm 0.3cm .0cm}]{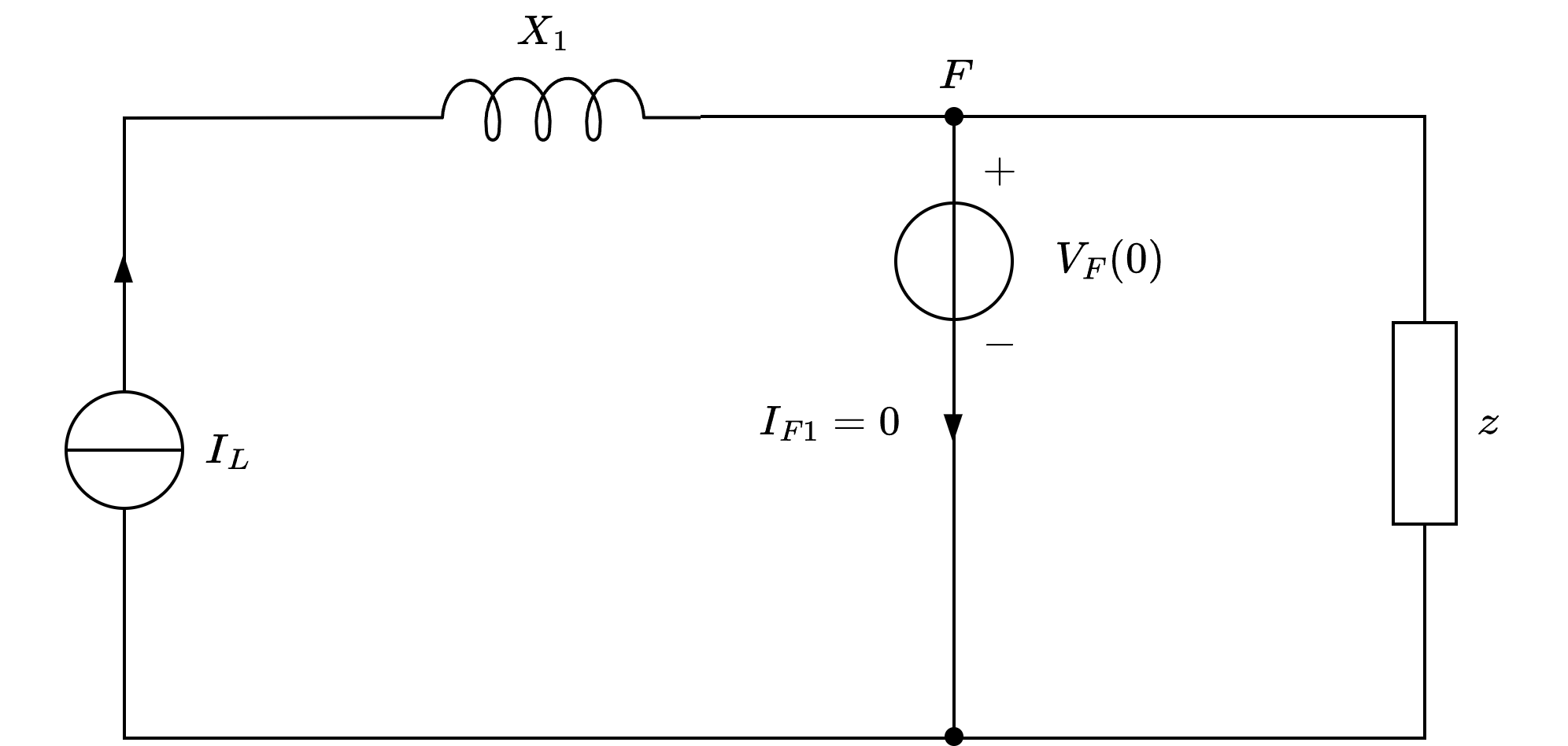}}
    \end{minipage} 
    \begin{minipage}{0.5\textwidth}
        \centering
        \scalebox{0.68}{\includegraphics[trim={0.3cm 0cm 0.8cm .0cm}]{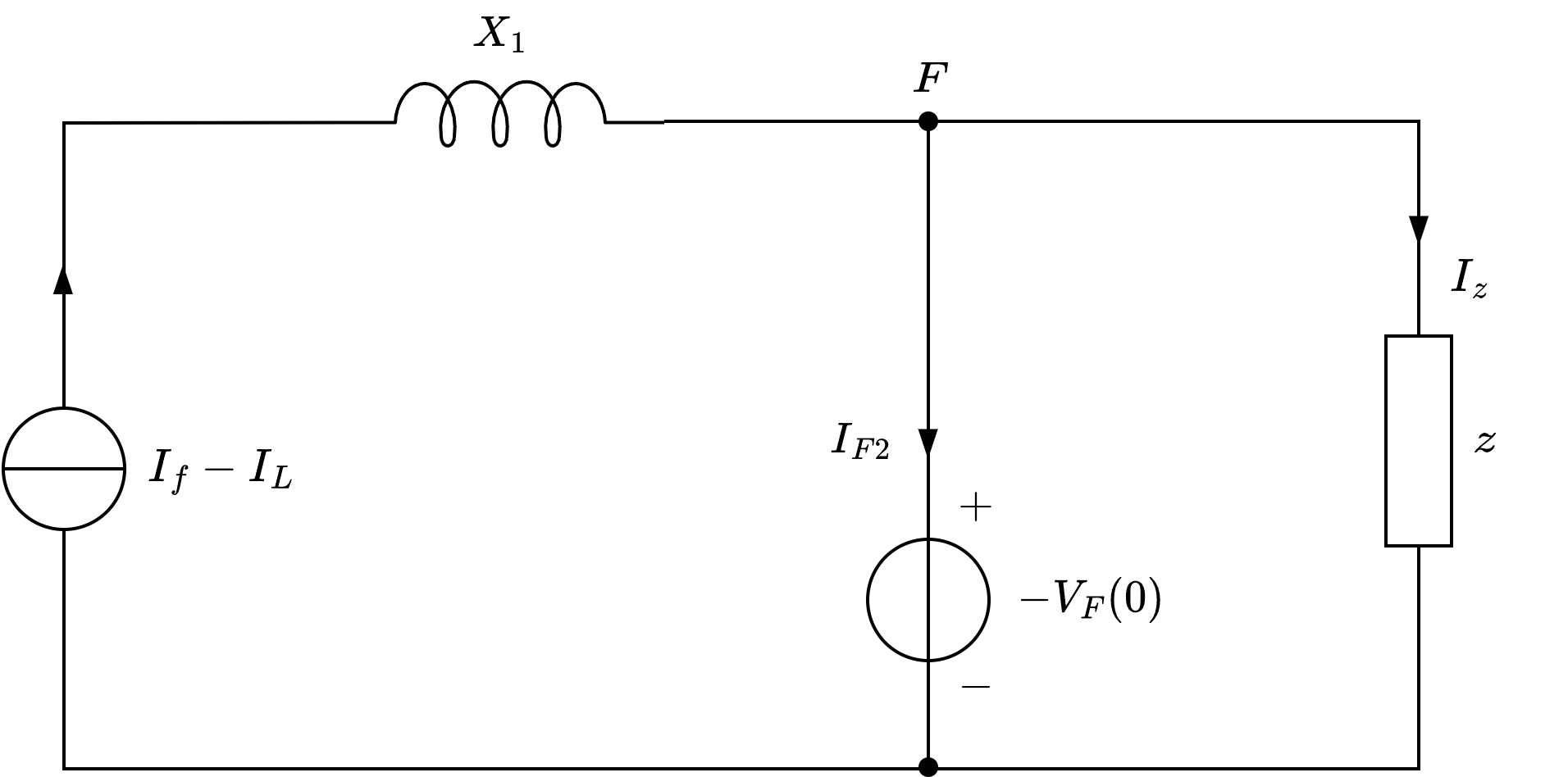}}
    \end{minipage} 
  \caption{\label{fig:VSC_SC}\textcolor{black}{Equivalent circuit to calculate the SCC from IBRs: (i) post-fault condition; (ii) pre-fault condition; (iii) `pure'-fault condition.}}
\end{figure}

    In Fig.~\ref{fig:VSC_SC}-(i), $I_f$ is a pre-defined parameter representing the short circuit injection from the IBR. Applying the superposition method, the post-fault system in Fig.~\ref{fig:VSC_SC}-(i) can be viewed as the combination of pre- and `pure'- fault conditions as shown in Fig.~\ref{fig:VSC_SC}-(ii) and (iii) respectively. In the pre-fault condition, $I_L$ is the load current from the IBR and $V_F(0)$ is the bus voltage at the fault location before the fault. A voltage source with the same magnitude is applied at this point without changing the system. Different from the calculation of SCC from SGs where all the SGs in the pure-fault condition are disregarded, the IBR in Fig.~\ref{fig:VSC_SC}-(iii) has to be considered as a current source with the output current being $I_f-I_L$ so that the superposition theorem holds. In order to validate the method to calculate the SCC from IBR, the SCC is derived from the post-fault system and the two sub-systems respectively. The SCC from the IBR $c\in \mathcal{C}$ can be derived from Fig.~\ref{fig:VSC_SC}-(i) as:
\begin{align}
\label{I_SC_VSC1}
    I_{F} = I_{fc}.    
\end{align}

It can also be derived as the linear combination of $I_{{F1}}$ and $I_{{F2}}$, i.e.:
\begin{align}
\label{I_SC_VSC2}
    I_{F} = I_{{F1}} + I_{{F2}}
\end{align}
where $ I_{{F1}}$ is zero and $ I_{{F2}}$ is computed by applying KCL in Fig.~\ref{fig:VSC_SC}-(iii):
\begin{align}
\label{I_SC2_VSC}
     I_{{F2}} = (I_{fc}-I_{L})-I_z.
    \end{align}
The load current in the `pure'-fault condition $I_z$ is derived from the pre-fault condition as follows:
\begin{align}
\label{Iz}
    I_z = \frac{-V_F(0)}{z} = -I_L.
\end{align}
Combining \eqref{I_SC_VSC2}, \eqref{I_SC2_VSC} and \eqref{Iz} gives \eqref{I_SC_VSC1}, which justifies the modified superposition method to calculate the SCC from IBRs. In a general multi-machine system with both SGs and IBRs, the SCC can be thus calculated by combining two `pure'-fault conditions, which modifies \eqref{IF_SG} as:
    \begin{align}
        -V_F(0) = \sum_{c\in \mathcal{C}} Z_{F\Phi(c)}(I_{fc}-I_{Lc})+Z_{FF}I_{F}.
    \end{align}

\bibliographystyle{IEEEtran}
\bibliography{bibliography}

\end{document}